\documentclass[12pt]{article}

\usepackage[utf8]{inputenc}
\usepackage[T1]{fontenc}
\usepackage{lmodern}
\usepackage[english]{babel}
\usepackage{amsmath, amssymb}
\usepackage{graphicx}
\usepackage{float}
\usepackage{subcaption} 
\usepackage{ragged2e}
\usepackage{hyperref}
\usepackage[noabbrev]{cleveref} 
\usepackage{cite}
\usepackage{geometry}
\usepackage{booktabs}  
\usepackage{siunitx}   
\title{Noncommutative Bianchi I and III Cosmology Models:\\ Radiation Era Dynamics and \(\gamma\) Estimation}

\author{
	G. Oliveira-Neto\thanks{Email: \texttt{gilneto@fisica.ufjf.br}} \\ 
	\small Departamento de Física, ICE, Universidade Federal de Juiz de Fora \\
	\and
	Y. Soncco Apaza\thanks{Email: \texttt{sonccoapaza.yuri@estudante.ufjf.br}} \\
	\small Departamento de Física, ICE, Universidade Federal de Juiz de Fora
}

\date{\today}

\begin{document}

\maketitle

\begin{abstract}
Understanding the early evolution of the universe requires models that incorporate possible quantum and anisotropic effects in its dynamics. In this work, we analyze the dynamical evolution of locally rotationally symmetric anisotropic cosmological models of Bianchi type I (flat curvature) and Bianchi type III (open curvature) within a noncommutative phase space framework characterized by a deformation parameter \(\gamma\). Using a Hamiltonian formulation based on Schutz’s formalism for a perfect radiation fluid, we introduce noncommutative Poisson brackets that allow for geometric corrections to commutative dynamics. The resulting equations are solved numerically under various initial conditions, enabling the study of the impact of \(\gamma\) and the energy density \(C\) on the universe’s expansion and anisotropy evolution. The results show that \(\gamma < 0\) enhances expansion and favors isotropization, while \(\gamma > 0\) tends to slow expansion and preserve residual anisotropy, especially in the open curvature model. It is estimated that the influence of non-commutativity was significant during the early stages of the universe, decreasing toward the present time, suggesting that this approach could serve as an effective alternative to the cosmological constant in describing the evolution of the early universe.
\end{abstract}

\vspace{1em}
\vspace{1.5em}

\section{\label{sec:level0} Introduction}  
Understanding the dynamics of the universe in its earliest stages remains one of the main challenges of modern cosmology. It is postulated that, during its initial moments when the energy density and spacetime curvature were extreme, the gravitational interaction was quantized \cite{rovelli2004quantum}. In this quantum-gravitational regime, the geometry of the universe could not be described by classical general relativity but instead behaved as a highly fluctuating structure, resembling a resonant quantum foam, where multiple spatial configurations coexist in superposition \cite{wheeler1964geometrodynamics}. This concept implies that the universe may have begun in an inhomogeneous and anisotropic state, far from the symmetric and uniform picture we observe today. However, over time, and driven by mechanisms such as accelerated expansion, this primordial universe would have evolved toward a simpler configuration: homogeneous and isotropic on large scales, as described by the standard Friedmann-Robertson-Walker (FRW) cosmological model \cite{mukhanov2005physical, weinberg2008cosmology}.

In this context, studies on the process of isotropization and how an initially anisotropic universe can evolve into an isotropic one have been widely developed with the aim of reconstructing the history of the early universe and exploring possible imprints of its quantum origin. From a classical perspective, it has been shown that homogeneous anisotropic models of Bianchi type tend to isotropic solutions in the presence of a positive cosmological constant \cite{wald1983asymptotic} or in the presence of a reduced relativistic gas (RRG) \cite{shapiro}. The same process of isotropization also takes place in Kantowski-Sachs cosmological models filled with a phantom perfect fluid \cite{gil10} or with a running cosmological constant \cite{gil11}. Moreover, dynamical analyses of these models have identified various conditions and mechanisms, such as cosmic viscosity, the presence of a radiation fluid, interactions with scalar fields, or even the action of dark energy, that can favor such evolution toward isotropy \cite{barrow2001isotropy, wainwright1997dynamical, heinzle2005matter}. 
In the specific case of Bianchi I cosmological models, we may mention several works dealing with the above-mentioned and other aspects of these models. In Ref. \cite{banerjee}, the authors investigated how the presence of a viscous fluid influences the isotropization of a Bianchi I model.
A nonminimal extension of the Einstein-Maxwell equations was coupled to Bianchi I models in the presence of a magnetic field in Ref \cite{balakin}. The authors obtained several exact solutions of the nonminimal system, including those that described isotropization, inflation, and an isotropic de Sitter solution characterized by a ‘screening’ of the magnetic field as a consequence of the nonminimal coupling. In Ref. \cite{canfora}, the authors considered the effects of the presence of Skyrmes and a positive cosmological constant in two anisotropic cosmologies. They studied Kantowski-Sachs and Bianchi-I universes and determined the conditions for the late time isotropization of both universes. The authors discussed in Ref. \cite{alexander}, the possible description of the big bang–big crunch cosmological
singularities crossing in a Bianchi I universe filled with minimally and conformally coupled massless scalar fields. In Ref. \cite{casadio} the authors studied the effects of a spatially homogenous magnetic field in Bianchi I cosmological models. They considered three models filled with a spatially homogeneous magnetic field (SHMF) and different matter contents. In the first one, there was just the SHMF, in the second one, an extra dust fluid, and in the third, an extra massless scalar field (stiff matter). They paid special attention to the identification of the initial and final states of these models and the transition between them.
In the quantum context, research developed within loop quantum cosmology has shown that specific anisotropic models, such as Bianchi I, can undergo isotropization as a natural result of the evolution of the universe in quantum-corrected geometries without the need to impose additional external conditions \cite{ashtekar2009loop, bojowald2008loop}. There are also works that studied Bianchi I cosmological models, at the quantum level, using the formalism of quantum cosmology \cite{eath,alvarenga,socorro,socorro1,socorro2}.

The Bianchi III model, which represents an anisotropic universe with negative spatial curvature, has been extensively investigated in both classical contexts and modified gravity theories. It has been employed to study isotropization mechanisms in anisotropic cosmologies \cite{byland1998evolution}, the dynamics of dark energy in general relativity \cite{pradhan2011dark}, and alternative formulations such as Lyra geometry with quadratic equations of state \cite{mollah2018bianchi}. Within modified gravity theories, the Bianchi III model has been analyzed in $f(R, T)$ gravity, exhibiting accelerated expansion and non-singular behaviors \cite{sahoo2016anisotropic}. Furthermore, quantum extensions of the model have been explored, for instance, within Hořava–Lifshitz gravity and in Bianchi III minisuperspace models coupled to electromagnetic fields, where both canonical classical quantization and Wheeler–DeWitt approaches have been employed \cite{christodoulakis2012classical}. These studies illustrate the use of the Bianchi III metric as a framework for examining anisotropies, isotropization, and cosmic acceleration within different theoretical approaches. 

Beyond these classical and modified gravity frameworks, another promising approach involves spacetime noncommutativity, which was first introduced by Snyder in the 1940s as an attempt to regularize divergences in quantum field theory without breaking Lorentz symmetry \cite{snyder1947quantized}. In his proposal, spacetime coordinates ceased to commute, giving rise to a modified geometric structure at tiny scales. Although initially overlooked, this idea experienced a significant resurgence due to developments in string theory, membrane theory, and M-theory, where noncommutativity naturally emerges in specific physical regimes \cite{seiberg1999string, connes1994noncommutative, douglas2001noncommutative}. In this new context, spacetime noncommutativity has become the subject of intense investigation, particularly in cosmology. Various studies have shown that noncommutative (NC) effects can profoundly alter the dynamics of the early universe \cite{barbosa2004noncommutative,bastos2008phase,socorro0,gil,gil1,gil2,gil3,gil4,gil5,gil6,gil7,gil8,gil9}, even contributing to the elimination of singularities and modifying the expansion process. This suggests that noncommutativity played an active role throughout all stages of cosmic evolution. One of the primary motivations for incorporating this structure into classical cosmological models is the possibility that residual observable effects may persist to the present day.

Consequently, a growing line of research in cosmology involves extending anisotropic models by incorporating NC structures in phase space. This proposal, motivated by advances in string theory and quantum gravity, introduces a parameter \(\gamma\) that deforms the Poisson algebra between dynamical variables. These modifications allow for the exploration of alternative evolutionary trajectories and have been associated with mechanisms that could suppress anisotropies, induce accelerated expansion, or eliminate singularities \cite{barbosa2004noncommutative,bastos2008phase}. However, the specific study of the Bianchi I (flat curvature) and Bianchi III (negative curvature) models during the radiation-dominated era within this NC framework remains limited. We may mention Refs. \cite{socorro0}, \cite{gil8} and \cite{gil9}, where NC Bianchi I models are studied. Moreover, to the best of our knowledge, the noncommutative (NC) deformation of Bianchi III cosmologies has not yet been explored, providing a motivation for its consideration in the present work.

Therefore, this study aims to analyze the dynamical evolution of the locally rotationally symmetric (LRS) NC Bianchi I and Bianchi III cosmological models, coupled to a perfect radiation fluid, investigating the effects of the NC parameter \(\gamma\) on the expansion and isotropization processes. Our main motivation to choose the LRS version of these spacetimes is the similarities between the metrics. As we shall see later, these similarities will allow us to compare some results coming from Bianchi I and III cosmological models in a simple way. 
As a matter of completeness, we mention that the LRS version of Bianchi I and III lets the metric of these spacetimes be very similar to the metric of the Kantowski–Sachs spacetime (positive curvature) \cite{kantowski1966some}. Unfortunately, as we shall see later, the Kantowski–Sachs model will not be studied in the present work because it does not produce expanding solutions under the conditions considered here.
The parameter $\gamma$ is introduced due to a deformed Poisson bracket algebra. More precisely, it appears because we introduce four nontrivial Poisson brackets between all geometrical variables and the matter variables of the model. We associate the same NC parameter $\gamma$ with all nontrivial Poisson brackets. 
An important difference between the present paper and the Refs. \cite{socorro0}, \cite{gil8} and \cite{gil9}, is that here we use the LRS form of the Bianchi I metric.
Also, we may say that the present model is more noncommutative than the one considered in Ref. \cite{gil8}. This is the case because the authors considered only two nontrivial Poisson brackets between the geometrical variables and the matter variables of the model. The content of matter in the model of Ref. \cite{gil8} is the same as in the present model. On the other hand, the authors in Ref. \cite{gil9} introduced six nontrivial Poisson brackets and in Ref. \cite{socorro0} four nontrivial Poisson brackets, all just between the geometrical variables of the model. The model in Ref. \cite{gil9} differs from the present model and from the ones in Refs. \cite{socorro0} and \cite{gil8}, because the content of matter there was given by an RRG. Another important difference between the present model and the one in Ref. \cite{gil9} is the use of the symplectic formalism to introduce the deformed Poisson bracket algebra. Finally, in Ref. \cite{socorro0}, the authors solved the dynamical equations in the WKB approximation, while here we solve the complete equations numerically. Besides, they considered a cosmological constant.
Additionally, here a quantitative estimation of \(\gamma\) is proposed on the basis of the assumption that the universe has evolved toward a homogeneous and isotropic state at present. This work seeks to evaluate whether noncommutativity may have played a significant role in the early universe and whether its influence can be distinguished from that of commutative models. Furthermore, a systematic comparison was made between the flat (Bianchi I) and open (Bianchi III) models, exploring their differences in terms of sensitivity to the NC deformation. This approach not only enriches the theoretical framework of anisotropic cosmology but may also open new avenues for connecting predictions from NC models with current or future cosmological observations.

Finally, the structure of this work is organized as follows. In Section 2, the Bianchi I and Bianchi III cosmological models coupled to a perfect fluid are introduced, and the superhamiltonian of the models is derived. In Section 3, we define the NC Poisson brackets through the parameter \(\gamma\), and the equations of motion are obtained in terms of new commutative variables. In Section 4, we consider the case of a radiation fluid. We numerically solve the system of dynamical equations and obtain the behaviors of the scale factor and the anisotropy parameter. We present a detailed study of their various behaviors and a comparison with the commutative case. In Section 5, the parameter \(\gamma\) is estimated from observational data, assuming a homogeneous and isotropic universe. Finally, in Section 6, we discuss and summarize the most relevant results of the paper.

\section{\label{sec:level1}  The Bianchi models for any perfect fluid}

The LRS line element, which encompasses the spacetimes of Bianchi I (BI), Bianchi III (BIII), and Kantowski-Sachs (KS) \cite{kantowski1966some, demianski1979physics, lorenz1982kantowski, akarsu, camci2016new}, allows a unified description of such homogeneous and anisotropic universes. Inspired by Misner’s parametrization \cite{misner1972minisuperspace}, this LRS line element can be expressed as \cite{gil10},
\begin{equation}\label{metric}
    \mathrm{d}s^2 = -N(t)^2 \mathrm{d}t^2 + a(t)^2 e^{-\beta(t)} \mathrm{d}r^2 + a(t)^2 e^{\beta(t)}\left( \mathrm{d}\theta^2 + f(\theta)^2 \mathrm{d}\phi^2\right).
\end{equation}
\noindent Here, $a(t)$ represents the isotropic scale factor, $\beta(t)$ is the parameter describing anisotropy, $N(t)$ is the lapse function, $f(\theta)$ takes the specific forms: $f(\theta) = \theta, \, \sin(\theta), \, \sinh(\theta)$, respectively, for the following spacetimes BI, KS, BIII, and we use the
natural unit system, where $c = 8\pi G = 1$. In the metric \(\eqref{metric}\), the anisotropy depends on the parameter $\beta(t)$, which produces differences in the expansion rates along different spatial directions. It is expected that $\beta(t) \to \text{constant or zero}$ at late times, implying that the universe, although initially anisotropic, evolves toward an isotropic FRW-like state. It is known in the literature that the Bianchi I and the Kantowski-Sachs spacetimes tend to the FRW spacetime at late times. However, the Bianchi III spacetime does not reach an exact FRW spacetime at late times. However, it becomes very similar to an FRW spacetime for a period of its evolution \cite{ellis}. Here, we are going to study the conditions in which the Bianchi III spacetime becomes very similar to an FRW spacetime.

The energy content of the universe can be represented by a perfect fluid. The four-velocity of this fluid, typically expressed as $U^{\mu}=N(t)\delta_{0}^{\mu}$, defines a comoving frame where matter exhibits no spatial motion and evolves only in time. This assumption significantly simplifies the calculations and provides a convenient basis for defining the energy-momentum tensor as
\begin{equation}\label{EnergyTensor}
    T_{\mu\nu} = \left(\rho + p \right) U_{\mu} U_{\nu} + p g_{\mu\nu}.
\end{equation}
\noindent Where $\rho$ represents the energy density, $p$ the pressure, and $g_{\mu\nu}$ is the metric tensor. 
The equation of state of the fluid is given by  
\begin{equation}\label{StateFluid}
    p = \alpha \rho, 
\end{equation}
\noindent where $\alpha$ is a constant that characterizes the type of fluid.

In this work, the Hamiltonian of a perfect fluid is determined using Schutz’s variational formalism \cite{schutz1970perfect}. In this approach, the four-velocity ($U_{\nu}$) of the fluid is expressed in terms of six thermodynamic potentials ($\lambda$,  $\varphi$, $\eta$, $\tau$, $\vartheta$, $s$) as follows,
\begin{equation}\label{CuadriVelocity}
    U_{\mu} = \frac{1}{\lambda} \left( \varphi_{,\mu} + \eta \tau_{,\mu} + \vartheta s_{,\mu} \right)
\end{equation}
\noindent where $\lambda$ is the specific enthalpy, $s$ is the specific entropy, $\eta$ and $\tau$ are associated with the vorticity of the fluid, while $\varphi$ and $\vartheta$ lack a clear physical interpretation in this context. Additionally, the four-velocity must satisfy the normalization condition, 
\begin{equation}\label{normalization}
    U^{\mu}U_{\mu}=-1.
\end{equation}

For these models, which do not assume vorticity, the potentials $\varphi$ and $\vartheta$ vanish. In a comoving coordinate system with the perfect fluid, the expression for $U_{\mu}$ simplifies, and the specific enthalpy $\lambda$ takes the form
\begin{equation}\label{Entalphy}
    \lambda = N^{-1} \left(\dot{\varphi} + \vartheta \dot{s}\right),
\end{equation}
where the dot means derivative with respect to the time $t$.

Considering the equation of state \eqref{StateFluid}, it can be shown by thermodynamic considerations that the pressure in this formalism has the form \cite{alvarenga2002quantum},
\begin{equation}\label{pressure}
    p = \alpha \left( \frac{\lambda}{\alpha + 1} \right)^{1 + 1/ \alpha} e^{-s/\alpha}.
\end{equation}
\noindent This formulation links the thermodynamic properties of the fluid with the dynamical variables in Schutz’s formalism. %

The action of the theory will be determined by the sum of the Einstein-Hilbert term and the action associated with the perfect fluid, and the total action can be written as
\begin{equation}\label{action}
    S = \int \left(\frac{1}{2} R + \mathcal{L}_{\mathrm{M}} \right) \sqrt{-g} \, \mathrm{d}^4 x
\end{equation}
\noindent where  $g$  is the determinant of the metric tensor,  $R$  is the Ricci scalar and $\mathcal{L}_{\mathrm{M}}$ is the Lagrangian density of matter, which is equal to the fluid pressure $p$. 

By computing the Ricci scalar R from the metric \eqref{metric} and substituting $p$ Eq. \eqref{pressure}, the action takes the explicit form
\begin{equation}\label{actionComputed}
    S = \int \left[  a N \kappa e^{-\frac{\beta}{2}} -\frac{a^2 \dot{a} \dot{\beta} e^{\frac{\beta}{2}}}{N}-\frac{3 a \dot{a}^2 e^{\frac{\beta}{2}}}{N}+\frac{a^3 \dot{\beta}^2 e^{\frac{\beta}{2}}}{4 N} + \alpha  a^3 N e^{\frac{\beta}{2}-\frac{s}{\alpha }} \left(\frac{\vartheta \dot{s}+\dot{\varphi}}{N(\alpha + 1)}\right)^{\frac{1}{\alpha }+1}\right] \mathrm{d} t
\end{equation}
\noindent where $\frac{f{\prime}{\prime}(\theta)}{f(\theta)} = -\kappa$ , with $\kappa = 0$ corresponding to the BI model,  $\kappa = -1$ to the BIII model, and $\kappa = 1$ to the KS model \cite{demianski1979physics, hawking2010general}.

The Lagrangian density can be obtained from the action of the model Eq. \eqref{actionComputed} and using it, one may write the corresponding superhamiltonian, using the geometrodynamic formulation of general relativity \cite{wheeler1964geometrodynamics},
\begin{equation}\label{SuperHamiltonian}
    N H = N e^{-\frac{\beta}{2}}\left(-\kappa a -\dfrac{P_{a}^{2}}{16a}-\dfrac{P_{a}P_{\beta }}{4a^{2}}+\dfrac{3P_{\beta }^{2}}{4a^{3}}\right) + N a^{3}e^{-\frac{\beta \alpha}{2}+s} P_{\varphi}^{\alpha +1}
\end{equation}
where $P_{a} = - N^{-1}a e^{\beta/2} \left(6 \dot{a} + a \dot{\beta}\right)$, $P_{\beta} = (2 N)^{-1}a^2 e^{\beta/2} \left(a \dot{\beta} - 2 \dot{a} \right)$, $P_{s} = \vartheta P_{\varphi}$, $P_{\vartheta} = P_{M} = 0$ and $P_{\varphi} = a^3 e^{\frac{\beta}{2}-\frac{s}{\alpha }} \left(\frac{\vartheta \dot{s}+\dot{\varphi}}{\alpha  M+M}\right)^{1/\alpha }$, are the canonically conjugate momenta of the dynamical variables of the model. The superhamiltonian \eqref{SuperHamiltonian} can be simplified by performing the following canonical transformations \cite{alvarenga2002quantum, lapchinskii1977quantum}, 
\begin{equation}\label{CanonicalTransportation}
    T = -P_{s}\,e^{-s}\,P_{\varphi}^{-(\alpha+1)} \mbox{,\,\,\,\,\, } P_{T} = P_{\varphi}^{\alpha + 1}\,e^{s} \mbox{,\,\,\,\,\,} \overline{\varphi} = \varphi - (\alpha -1)\frac{P_{s}}{P_{\varphi}} \mbox{,\,\,\,\,\,} \overline{P}_{\varphi} = P_{\varphi}.
\end{equation}
\noindent Using the transformations \eqref{CanonicalTransportation} in the matter term of the superhamiltonian \eqref{SuperHamiltonian}, it becomes a simpler expression, 
\begin{equation}\label{SuperHamiltonianMatterOnly}
     H =  e^{-\beta/2}\left(-\kappa a -\dfrac{P_{a}^{2}}{16a}-\dfrac{P_{a}P_{\beta }}{4a^{2}}+\dfrac{3P_{\beta }^{2}}{4a^{3}}\right) +  a ^{-3\alpha} e^{-\beta\alpha/2} P_{T}.
\end{equation}
\noindent Here, $P_{T}$ is the only remaining canonical variable associated with matter \cite{alvarenga2002quantum}. For simplicity, here we use the gauge $N=1$.

\section{\label{sec:level3}  The Noncommutative Bianchi models for any perfect fluid} 

We consider that the NC Hamiltonian has the same form as the commutative Hamiltonian \eqref{SuperHamiltonianMatterOnly}. It can be rewritten simply by placing the $nc$ index in the commutative variables, which we call NC variables, to differentiate it from the old variables. As we write below,
\begin{equation}\label{NCSuperHamiltonianMatter}
     H_{nc} =  e^{-\beta_{nc}/2}\left(-\kappa a_{nc} -\dfrac{P_{a_{nc}}^{2}}{16a_{nc}}-\dfrac{P_{a_{nc}}P_{\beta_{nc} }}{4a_{nc}^{2}}+\dfrac{3P_{\beta_{nc} }^{2}}{4a_{nc}^{3}}\right) +  a_{nc}^{-3\alpha} e^{-\beta_{nc}\alpha/2} P_{T_{nc}}.
\end{equation}
\noindent Here, $P_{a_{nc}}, \, P_{\beta_{nc}}, \, P_{T_{nc}}$ are, respectively, the canonically conjugated momenta to the coordinates $a_{nc},\, \beta_{nc},\, T_{nc}$ in the NC model.

The NC model is particularly relevant for exploring the early universe, where quantum gravity effects could be significant \cite{douglas2001noncommutative}. In NC cosmology, it is assumed that some of the classical phase space variables do not commute, which introduces a modified algebra in this space \cite{bastos2008phase, barbosa2004noncommutative,  gil1, socorro2024non}. For the present model, we introduce the following Poisson brackets (PBs): 
\begin{eqnarray}\label{PoissonBracket a}
    & &\left\lbrace a_{nc},~T_{nc} \right\rbrace =\left\lbrace \beta_{nc},~T_{nc} \right\rbrace =\left\lbrace P_{a_{nc}},~P_{T_{nc}} \right\rbrace =\left\lbrace P_{\beta_{nc}},~P_{T_{nc}} \right\rbrace=0 \\ \nonumber
    && \left\lbrace a_{nc},\, \beta_{nc} \right\rbrace = \left\lbrace a_{nc},\, P_{\beta_{nc}} \right\rbrace = \left\lbrace \beta_{nc},\, P_{a_{nc}} \right\rbrace = \left\lbrace P_{a_{nc}},\, P_{\beta_{nc}} \right\rbrace = 0
\end{eqnarray}
\begin{eqnarray} \label{PoissonBracket  b}
     & &\left\lbrace a_{nc},~P_{a_{nc}} \right\rbrace =\left\lbrace \beta_{nc},~P_{\beta_{nc}} \right\rbrace =\left\lbrace T_{nc},~P_{T_{nc}} \right\rbrace =1
\end{eqnarray}
\begin{eqnarray} \label{PoissonBracket c}
     & & \left\lbrace a_{nc},~P_{T_{nc}} \right\rbrace =\left\lbrace \beta_{nc},~P_{T_{nc}} \right\rbrace =\left\lbrace T_{nc},~P_{a_{nc}} \right\rbrace =\left\lbrace T_{nc},~P_{\beta_{nc}} \right\rbrace =\gamma.
\end{eqnarray}
\noindent Here, $\gamma$ is the NC parameter that should be a small quantity in our current universe. All effects resulting from noncommutativity are concentrated in this parameter. In a system where phase space variables are NC, the expansion of the universe is influenced by the NC structure.

Now, instead of working directly with the physical NC variables of the Hamiltonian \eqref{NCSuperHamiltonianMatter}, as a matter of simplicity, we introduce new auxiliary canonical variables $(a_{c},\, \beta_{c},\, T_{c},\, P_{a_{c}}, \, P_{\beta_{c}}, \, P_{T_{c}})$, defined as 
\begin{equation}\label{EquivalencesNC}
    	\begin{aligned}
    	a_{nc} & \to a_{c} + (\gamma/2) T_{c}, ~~ P_{a_{nc}}  \to P_{a_{c}} + (\gamma/2) P_{T_{c}},\\
        \\
    	T_{nc} & \to T_{c} + (\gamma/2) a_{c}, ~~ P_{T_{nc}} \to P_{T_{c}} + (\gamma/2) P_{a_{c}},\\
        \\
    	\beta_{nc} & \to \beta_{c}, ~~ P_{\beta_{nc}} \to P_{\beta_{c}}.
    	\end{aligned}
\end{equation}
\noindent Here, $(a_{c},\, \beta_{c},\, T_{c})$ and $(P_{a_{c}}, \, P_{\beta_{c}}, \, P_{T_{c}})$ are new coordinates and momenta in a commutative phase space. They satisfy the usual commutative Poisson brackets: $\left\lbrace a_{c},~P_{a_{c}} \right\rbrace =\left\lbrace \beta_{c},\,~P_{\beta_{c}} \right\rbrace =\left\lbrace T_{c},\,~P_{T_{c}} \right\rbrace =1$ and all other Poisson brackets are zero. It is possible to show that, if one introduces the new expressions of the NC variables, given in terms of the commutative variables and $\gamma$ Eqs. \eqref{EquivalencesNC}, they fully satisfy the Poisson brackets \eqref{PoissonBracket a}-\eqref{PoissonBracket c}, up to the first order in $\gamma$. These transformations Eqs. \eqref{EquivalencesNC} are also known as Bopp shifts \cite{das2009non}. This technique of Bopp shift transformations is particularly significant in cosmology, as it provides a method to incorporate noncommutativity into classical systems by modifying the phase space coordinates \cite{barbosa2004noncommutative, socorro2024non}.

With the aid of the transformations Eqs. \eqref{EquivalencesNC}, we can rewrite the Hamiltonian \eqref{NCSuperHamiltonianMatter} in terms of the new coordinates and momenta, up to the first order in $\gamma$, which gives us
\begin{equation}\label{NCSuperHamiltonianMatterPertuvative}
    \begin{aligned}
     H_{c} =  e^{-\frac{\beta_{c}}{2}}\left(-\kappa\left(a_{c} + \frac{\gamma}{2}T_{c}\right) -\dfrac{\left(P_{a_{c}} + \frac{\gamma}{2} P_{T_{c}}\right)^{2}}{16\left(a_{c} + \frac{\gamma}{2}T_{c}\right)}-\dfrac{\left(P_{a_{c}} + \frac{\gamma}{2} P_{T_{c}}\right)P_{\beta_{c} }}{4\left(a_{c} + \frac{\gamma}{2}T_{c}\right)^{2}}+\dfrac{3P_{\beta_{c} }^{2}}{4\left(a_{c} + \frac{\gamma}{2}T_{c}\right)^{3}}\right) \\
     +  \left(a_{c} + \frac{\gamma}{2}T_{c}\right)^{-3\alpha} e^{-\frac{1}{2}\beta_{c}\alpha} \left(P_{T_{c}} + \frac{\gamma}{2} P_{a_{c}}\right).
     \end{aligned}
\end{equation}
\noindent Here,  $H_{c}$  represents the NC Hamiltonian, expressed in terms of the auxiliary commutative variables denoted by the subscript $c$.%

The Hamiltonian formalism provides a framework that describes the time evolution of a system through Hamilton’s equations. Using the Hamiltonian $H_{c}$ \eqref{NCSuperHamiltonianMatterPertuvative}, the equations of motion for the BI, KS, and BIII models are given by:
\begin{equation}\label{MovmentEquation a}
    \begin{aligned}
     \dot{a}_{c} = \{a_{c},\, H_{c}\} = -\frac{e^{-\frac{\beta_{c}}{2}} P_{a_{c}}}{8 \left(a_{c}+\frac{\gamma}{2}T_{c}\right)}-\frac{e^{-\frac{\beta_{c}}{2}} P_{\beta_{c}}}{4 \left(a_{c}+\frac{\gamma}{2}T_{c}\right)^2}-\frac{\gamma  e^{-\frac{\beta_{c}}{2}} P_{T_{c}}}{16 \left(a_{c} +\frac{\gamma}{2} T_{c}\right)} \\
     +\frac{1}{2} \gamma  e^{-\frac{1}{2} \alpha \beta_{c}} \left(a_{c} +\frac{\gamma}{2}T_{c}\right)^{-3 \alpha },
     \end{aligned}
\end{equation}
\begin{equation}\label{MovmentEquation b}
    \begin{aligned}
     \dot{P}_{a_{c}} = \{P_{a_{c}},\, H_{c}\}= -\frac{e^{-\frac{\beta_{c}}{2}} P_{a_{c}} P_{\beta_{c}}}{2 \left(a_{c}+\frac{\gamma}{2}  T_{c}\right)^3}-\frac{\gamma  e^{-\frac{\beta_{c}}{2}} P_{a_{c}} P_{T_{c}}}{16 \left(a_{c}+\frac{\gamma}{2}  T_{c}\right)^2}-\frac{e^{-\frac{\beta_{c}}{2}} P_{a_{c}}^2}{16 \left(a_{c}+\frac{\gamma}{2} T_{c}\right)^2}\\
     +\frac{3}{2} \alpha  \gamma P_{a_{c}}  e^{-\frac{1}{2} \alpha \beta_{c}} \left(a_{c} +\frac{\gamma}{2} T_{c}\right)^{-3 \alpha -1} -\frac{\gamma  e^{-\frac{\beta_{c}}{2}} P_{\beta_{c}} P_{T_{c}}}{4 \left(a_{c} +\frac{\gamma}{2}  T_{c}\right)^3}\\
     +\frac{9 e^{-\frac{\beta_{c}}{2}} P_{\beta_{c}}^2}{4 \left(a_{c} +\frac{\gamma}{2}  T_{c}\right)^4}+ 3 \alpha P_{T_{c}} e^{-\frac{1}{2} \alpha  \beta_{c}} \left(a_{c} +\frac{\gamma}{2}  T_{c}\right)^{-3 \alpha -1}+ \kappa e^{-\frac{\beta_{c}}{2}},
     \end{aligned}
\end{equation}
\begin{equation}\label{MovmentEquation c}
    \begin{aligned}
     \dot{\beta_{c}} = \{\beta_{c},\, H_{c}\}= -\frac{e^{-\frac{\beta_{c}}{2}} P_{a_{c}}}{4 \left(a_{c} +\frac{\gamma}{2}  T_{c}\right)^2} +\frac{3 e^{-\frac{\beta_{c}}{2}} P_{\beta_{c}}}{2 \left(a_{c} +\frac{\gamma}{2}  T_{c}\right)^3}-\frac{\gamma  e^{-\frac{\beta_{c}}{2}} P_{T_{c}}}{8 \left(a_{c}+\frac{\gamma}{2} T_{c}\right)^2},
     \end{aligned}
\end{equation}
\begin{equation}\label{MovmentEquation d}
    \begin{aligned}
     \dot{P}_{\beta_{c}} = \{P_{\beta_{c}},\, H_{c}\} = -\frac{e^{-\frac{\beta_{c}}{2}} P_{a_{c}} P_{\beta_{c}}}{8 \left(a_{c}+\frac{\gamma}{2}  T_{c}\right)^2}-\frac{\gamma  e^{-\frac{\beta_{c}}{2}} P_{a_{c}} P_{T_{c}}}{32 \left(a_{c}+\frac{\gamma}{2}  T_{c}\right)}+\frac{1}{4} \alpha  \gamma P_{a_{c}} e^{-\frac{1}{2} \alpha  \beta_{c}} \left(a_{c} + \frac{\gamma}{2} T_{c}\right)^{-3 \alpha }\\
     -\frac{e^{-\frac{\beta_{c}}{2}} P_{a_{c}}^2}{32 \left(a_{c}+\frac{\gamma}{2}  T_{c}\right)}-\frac{\gamma  e^{-\frac{\beta_{c}}{2}} P_{\beta_{c}} P_{T_{c}}}{16 \left(a_{c} +\frac{\gamma}{2}  T_{c}\right)^2} + \frac{3 e^{-\frac{\beta_{c}}{2}} P_{\beta_{c}}^2}{8 \left(a_{c}+\frac{\gamma}{2}  T_{c}\right)^3}\\
     +\frac{1}{2} \alpha P_{T_{c}} e^{-\frac{1}{2} \alpha  \beta_{c}} \left(a_{c}+\frac{\gamma}{2}  T_{c}\right)^{-3 \alpha }-\frac{1}{2} \kappa a_{c} e^{-\frac{\beta_{c}}{2}}-\frac{1}{4} \kappa \gamma  e^{-\frac{\beta_{c}}{2}} T_{c},
     \end{aligned}
\end{equation}
\begin{equation}\label{MovmentEquation e}
    \begin{aligned}
     \dot{T}_{c} = \{T_{c},\, H_{c}\} = -\frac{\gamma  e^{-\frac{\beta_{c}}{2}} P_{a_{c}}}{16 \left(a_{c}+\frac{\gamma}{2}  T_{c}\right)}-\frac{\gamma  e^{-\frac{\beta_{c}}{2}} P_{\beta_{c}}}{8 \left(a_{c}+\frac{\gamma}{2} T_{c}\right)^2}+e^{-\frac{1}{2} \alpha  \beta_{c}} \left(a_{c}+\frac{\gamma}{2}  T_{c}\right)^{-3 \alpha },
     \end{aligned}
\end{equation}
\begin{equation}\label{MovmentEquation f}
    \begin{aligned}
     \dot{P}_{T_{c}} =\{P_{T_{c}},\, H_{c}\} = -\frac{\gamma  e^{-\frac{\beta_{c}}{2}} P_{a_{c}} P_{\beta_{c}}}{4 \left(a_{c}+\frac{\gamma}{2}  T_{c}\right)^3}-\frac{\gamma  e^{-\frac{\beta_{c}}{2}} P_{a_{c}}^2}{32 \left(a_{c}+\frac{\gamma}{2}  T_{c}\right)^2}+\frac{9 \gamma  e^{-\frac{\beta_{c}}{2}} P_{\beta_{c}}^2}{8 \left(a_{c}+\frac{\gamma}{2}  T_{c}\right)^4}\\
     +\frac{3}{2} \alpha  \gamma P_{T_{c}}  e^{-\frac{1}{2} \alpha  \beta_{c}} \left(a_{c}+\frac{\gamma}{2}  T_{c}\right)^{-3 \alpha -1}+\frac{1}{2} \kappa \gamma e^{-\frac{\beta_{c}}{2}}.
     \end{aligned}
\end{equation}

To analyze the evolution of the universe, our goal is to solve the equations corresponding to the variables \(a_c\), \(\beta_c\), and \(T_c\), which are related to the physical variables \(a_{nc}\) and \(\beta_{nc}\), as mentioned earlier. To achieve this, it is necessary to rewrite the equations of motion derived from the Hamiltonian \(H_c\), expressing them exclusively in terms of these variables and their time derivatives, thus eliminating explicit dependencies on the conjugate momenta. This approach simplifies the description of the dynamics of the system and makes it easier to analyze the impact of the corrections in \(\gamma\) on the evolution of the dynamic variables.

We observe that \(\dot{P}_{T_c}\) Eq. \eqref{MovmentEquation f} is proportional to \(\dot{P}_{a_c}\) Eq. \eqref{MovmentEquation b} with a factor of \(\gamma/2\). This implies \(\dot{P}_{T_c} = \frac{\gamma}{2} \dot{P}_{a_c}\). Integrating this equation, we obtain the following expression for the canonical momentum,
\begin{equation}\label{CanonicalMomentumPT}
    \begin{aligned}
        P_{T_{c}} = \frac{\gamma}{2} P_{a_{c}} + C.
     \end{aligned}
\end{equation}
\noindent where \(C\) is a positive integration constant that represents the energy density of the fluid. This constant is a quantity that is conserved during the evolution of the system and is determined by the initial conditions.

Also, if we choose  \eqref{MovmentEquation a}  and  \eqref{MovmentEquation c} from the equations of motion above and combine them, with the aid of Eq. \eqref{CanonicalMomentumPT}, we may obtain the following expressions for canonical momenta  $P_{a_{c}}$  and  $P_{\beta_{c}}$,  
\begin{equation}\label{CanonicalMomentumPa}
    \begin{aligned}
        P_{a_{c}}=-3 \gamma  e^{\frac{\beta_{c}}{2}} T_{c} \dot{a}_{c}-6 a_{c} e^{\frac{\beta_{c}}{2}} \dot{a}_{c}-\gamma  a_{c} e^{\frac{\beta_{c}}{2}} T_{c} \dot{\beta}_{c}-a_{c}^2 e^{\frac{\beta_{c}}{2}} \dot{\beta}_{c}\\
        +3 \gamma  e^{\frac{\beta_{c}}{2}-\frac{1}{2} \alpha  \beta_{c}} \left(a_{c}+\frac{1}{2} \gamma  T_{c}\right)^{1-3 \alpha }-\frac{C \gamma }{2},
     \end{aligned}
\end{equation}
\begin{equation}\label{CanonicalMomentumPb}
    \begin{aligned}
       P_{\beta_{c}} = -\gamma  a_{c} e^{\frac{\beta_{c}}{2}} T_{c} \dot{a}_{c}-a_{c}^2 e^{\frac{\beta_{c}}{2}} \dot{a}_{c}+\frac{3}{4} \gamma  a_{c}^2 e^{\frac{\beta_{c}}{2}} T_{c} \dot{\beta}_{c}+\frac{1}{2} a_{c}^3 e^{\frac{\beta_{c}}{2}} \dot{\beta}_{c}\\
       +\frac{1}{2} \gamma  e^{\frac{\beta_{c}}{2}-\frac{1}{2} \alpha  \beta_{c}} \left(a_{c}+\frac{1}{2} \gamma  T_{c}\right)^{2-3 \alpha }.
     \end{aligned}
\end{equation}
\noindent These equations express the canonical momenta \(P_{a_c}\) and \(P_{\beta_c}\) in terms of the variables \(a_c\), \(\beta_c\), \(T_c\), and their time derivatives.

Now, substituting the canonical momenta \(P_{T_c}\), \(P_{a_c}\), and \(P_{\beta_c}\) from equations \eqref{CanonicalMomentumPT}, \eqref{CanonicalMomentumPa}, and \eqref{CanonicalMomentumPb} into the time evolution of the matter-associated variable Eq. \eqref{MovmentEquation e}, we obtain the following expression for \(\dot{T}_c\),
\begin{equation}\label{dT}
    \begin{aligned}
       \dot{T}_{c} = \frac{1}{2} \gamma  \dot{a}_{c}+e^{-\frac{1}{2} \alpha  \beta_{c}} \left(a_{c}+\frac{1}{2} \gamma  T_{c}\right)^{-3 \alpha }.
     \end{aligned}
\end{equation}
\noindent This equation describes the variable \(\dot{T}_c\) in terms of \(a_c\), \(\dot{a}_c\), and \(T_c\), up to the first order in the NC parameter $\gamma$. %

Furthermore, it is necessary to derive, with respect to time, \(\dot{a}_c\) and \(\dot{\beta}_c\), as presented in equations \eqref{MovmentEquation a} and \eqref{MovmentEquation c}, respectively. This process allows for obtaining second-order differential equations, which are fundamental to fully describing the temporal evolution of the system's variables. Differential equations provide detailed information on how \(a_c\) and \(\beta_c\) evolve over time. Furthermore, deriving these expressions ensures a consistent physical description aligned with the system's initial conditions and constraints. These second derivatives are as follows:

\begin{equation}\label{dda}
    \begin{aligned}
       \ddot{a}_{c} &= -\frac{3}{4} \gamma  \dot{a}_{c} e^{-\frac{1}{2} \alpha  \beta_{c}} \left(a_{c}+\frac{1}{2} \gamma  T_{c}\right)^{-3 \alpha -1}+\frac{3}{2} \alpha  \gamma  \dot{a}_{c} e^{-\frac{1}{2} \alpha  \beta_{c}} \left(a_{c}+\frac{1}{2} \gamma  T_{c}\right)^{-3 \alpha -1} -\\
       &\frac{\dot{a}_{c}^2}{2 \left(a_{c}+\frac{1}{2} \gamma  T_{c}\right)}+\frac{1}{8} \gamma  e^{-\frac{1}{2} \alpha  \beta_{c}} \dot{\beta}_{c} \left(a_{c}+\frac{1}{2} \gamma  T_{c}\right)^{-3 \alpha }+\frac{1}{4} \alpha  \gamma  e^{-\frac{1}{2} \alpha  \beta_{c}} \dot{\beta}_{c} \left(a_{c}+\frac{1}{2} \gamma  T_{c}\right)^{-3 \alpha }\\
       &-\frac{1}{8} \dot{\beta}_{c}^2 \left(a_{c}+\frac{1}{2} \gamma  T_{c}\right)-\frac{1}{2} \alpha  C e^{-\frac{1}{2} \alpha  \beta_{c}-\frac{\beta_{c}}{2}} \left(a_{c}+\frac{1}{2} \gamma  T_{c}\right)^{-3 \alpha -2}-\frac{\kappa  e^{-\beta_{c}}}{8 \left(a_{c}+\frac{1}{2} \gamma  T_{c}\right)}\\
       &+\frac{\kappa  a_{c} e^{-\beta_{c}}}{8 \left(a_{c}+\frac{1}{2} \gamma  T_{c}\right)^2}+\frac{\gamma  \kappa  e^{-\beta_{c}} T_{c}}{16 \left(a_{c}+\frac{1}{2} \gamma  T_{c}\right)^2}
     \end{aligned}
\end{equation}
and 
\begin{equation}\label{ddb}
    \begin{aligned}
     \ddot{\beta}_{c} =  -\frac{3 \dot{a}_{c} \dot{\beta}_{c}}{a_{c}+\frac{1}{2} \gamma  T_{c}}+\frac{3}{2} \gamma  \dot{a}_{c} e^{-\frac{1}{2} \alpha  \beta_{c}} \left(a_{c}+\frac{1}{2} \gamma  T_{c}\right)^{-3 \alpha -2}-\frac{\kappa  e^{-\beta_{c}}}{4 \left(a_{c}+\frac{1}{2} \gamma  T_{c}\right)^2}-\frac{1}{2} \dot{\beta}_{c}^2 \\
     -\frac{5}{4} \gamma  e^{-\frac{1}{2} \alpha  \beta_{c}} \dot{\beta}_{c} \left(a_{c}+\frac{1}{2} \gamma  T_{c}\right)^{-3 \alpha -1}-\frac{3 \kappa  a_{c} e^{-\beta_{c}}}{4 \left(a_{c}+\frac{1}{2} \gamma  T_{c}\right)^3}-\frac{3 \gamma  \kappa  e^{-\beta_{c}} T_{c}}{8 \left(a_{c}+\frac{1}{2} \gamma  T_{c}\right)^3}
     \end{aligned}
\end{equation}
\noindent In this case, the derivatives of the canonical momenta $\dot{P}_{a_{c}}$ \eqref{MovmentEquation b}, $\dot{P}_{\beta_{c}}$ \eqref{MovmentEquation d}, $\dot{P}_{T_{c}}$ \eqref{MovmentEquation f} and the conjugate canonical momenta $P_{a_{c}}$ \eqref{CanonicalMomentumPa}, $P_{T_{c}}$ \eqref{CanonicalMomentumPT}, $P_{\beta_{c}}$ \eqref{CanonicalMomentumPb} were substituted. 

The Hamiltonian \(H_{c}\) plays a crucial role as a constraint in the system \((H_{c} = 0)\), ensuring that the dynamic trajectories satisfy the equations of motion along with the physical conditions established by the model. By substituting the conjugate canonical momenta \(P_{a_{c}} \eqref{CanonicalMomentumPa}, P_{T_{c}} \eqref{CanonicalMomentumPT}, P_{\beta_{c}}\)\eqref{CanonicalMomentumPb} into the expression \eqref{NCSuperHamiltonianMatterPertuvative}, we obtain
\begin{equation}\label{ConstrictionHc}
    \begin{aligned}
     0  = -e^{\frac{\beta_{c}}{2}} \dot{a}_{c} \dot{\beta}_{c} \left(a_{c}+\frac{1}{2} \gamma  T_{c}\right)^2-3 e^{\frac{\beta_{c}}{2}} \dot{a}_{c}^2 \left(a_{c}+\frac{1}{2} \gamma  T_{c}\right)+\frac{1}{4} e^{\frac{\beta_{c}}{2}} \dot{\beta}_{c}^2 \left(a_{c}+\frac{1}{2} \gamma  T_{c}\right)^3\\
     -\frac{1}{2} \gamma  e^{\frac{\beta_{c}}{2}-\frac{1}{2} \alpha  \beta_{c}} \dot{\beta}_{c} \left(a_{c}+\frac{1}{2} \gamma  T_{c}\right)^{2-3 \alpha }-3 \gamma  \dot{a}_{c} e^{\frac{\beta_{c}}{2}-\frac{1}{2} \alpha  \beta_{c}} \left(a_{c}+\frac{1}{2} \gamma  T_{c}\right)^{1-3 \alpha }\\
     +C e^{-\frac{1}{2} \alpha  \beta_{c}} \left(a_{c}+\frac{1}{2} \gamma  T_{c}\right)^{-3 \alpha }-\kappa  e^{-\frac{\beta_{c}}{2}} \left(a_{c}+\frac{1}{2} \gamma  T_{c}\right).
     \end{aligned}
\end{equation}
\noindent This expression illustrates the dynamics dictated by the constraints of the Hamiltonian system, where the geometry of the universe (\(\kappa\)) and the corrections due to the NC parameter \(\gamma\) are incorporated within \(H_{c}\).

\section{\label{sec:level 4} Numerical solution for a radiation-dominated anisotropic universe }

From now on, we restrict our attention to the case of a radiation perfect fluid where \(\alpha = 1/3\). The analytical solution of the coupled system of equations \eqref{dT}-\eqref{ConstrictionHc} is unfeasible due to the nonlinearity of the terms in its variables for \(\alpha = 1/3\). Therefore, we must solve it numerically. Furthermore, as mentioned above \(\gamma \ll 1\). Then, using this property, in terms of the equations where \(\gamma\) appears in the denominator, a Taylor expansion is performed up to the first order in \(\gamma\), which keeps all the \(\gamma\)'s in the numerator. With these modifications, the system of equations and the Hamiltonian constraint are rewritten in such a way that they contain only terms up to the first order in \(\gamma\), allowing for the determination of numerical solutions for the variables \(a_c(t)\), \(\beta_c(t)\), and \(T_c(t)\). These equations are given by:

\begin{equation}\label{approx_dda}
    \begin{aligned}
     \ddot{a}_{c} = -\frac{\dot{a}_{c}^2}{2 a_{c}}-\frac{\gamma  e^{-\frac{\beta_{c}}{6}} \dot{a}_{c}}{4 a_{c}^2}+\frac{\gamma  T_{c} \dot{a}_{c}^2}{4 a_{c}^2}+\frac{5 \gamma  e^{-\frac{\beta_{c}}{6}} \dot{\beta}_{c}}{24 a_{c}}-\frac{1}{8} a_{c} \dot{\beta}_{c}^2+\frac{C \gamma  e^{-\frac{1}{3} (2 \beta_{c})} T_{c}}{4 a_{c}^4}-\frac{C e^{-\frac{1}{3} (2 \beta_{c})}}{6 a_{c}^3}\\
     -\frac{1}{16} \gamma  T_{c} \dot{\beta}_{c}^2,
     \end{aligned}
\end{equation}

\begin{equation}\label{approx_ddb}
    \begin{aligned}
     \ddot{\beta}_{c} = -\frac{3 \dot{a}_{c} \dot{\beta}_{c}}{a_{c}}+\frac{3 \gamma  T_{c} \dot{a}_{c} \dot{\beta}_{c}}{2 a_{c}^2}+\frac{3 \gamma  e^{-\frac{\beta_{c}}{6}} \dot{a}_{c}}{2 a_{c}^3}-\frac{5 \gamma  e^{-\frac{\beta_{c}}{6}} \dot{\beta}_{c}}{4 a_{c}^2}-\frac{\kappa  e^{-\beta_{c}}}{a_{c}^2}+\frac{\gamma  \kappa  e^{-\beta_{c}} T_{c}}{a_{c}^3}-\frac{1}{2} \dot{\beta}_{c}^2,
     \end{aligned}
\end{equation}

\begin{equation}\label{approx_dT}
    \begin{aligned}
     \dot{T}_{c} = \frac{e^{-\frac{\beta_{c}}{6}}}{a_{c}}+\frac{1}{2} \gamma  \dot{a}_{c}-\frac{\gamma  e^{-\frac{\beta_{c}}{6}} T_{c}}{2 a_{c}^2},
     \end{aligned}
\end{equation}
and the constraint \((H_{c} = 0)\) is:
\begin{equation}\label{approx_Hc}
    \begin{aligned}
     0 = -a_{c}^2 e^{\frac{\beta_{c}}{2}} \dot{a}_{c} \dot{\beta}_{c}-\gamma  a_{c} e^{\frac{\beta_{c}}{2}} T_{c} \dot{a}_{c} \dot{\beta}_{c}-3 \gamma  e^{\frac{\beta_{c}}{3}} \dot{a}_{c}-\frac{3}{2} \gamma  e^{\frac{\beta_{c}}{2}} T_{c} \dot{a}_{c}^2-3 a_{c} e^{\frac{\beta_{c}}{2}} \dot{a}_{c}^2-\frac{1}{2} \gamma  a_{c} e^{\frac{\beta_{c}}{3}} \dot{\beta}_{c}\\
     +\frac{3}{8} \gamma  a_{c}^2 e^{\frac{\beta_{c}}{2}} T_{c} \dot{\beta}_{c}^2+\frac{1}{4} a_{c}^3 e^{\frac{\beta_{c}}{2}} \dot{\beta}_{c}^2-\frac{C \gamma  e^{-\frac{\beta_{c}}{6}} T_{c}}{2 a_{c}^2}+\frac{C e^{-\frac{\beta_{c}}{6}}}{a_{c}}-\kappa  a_{c} e^{-\frac{\beta_{c}}{2}}-\frac{1}{2} \gamma  \kappa  e^{-\frac{\beta_{c}}{2}} T_{c}.
     \end{aligned}
\end{equation}

This system of equations describes the temporal evolution of the commutative variables \(a_c(t)\), \(\beta_c(t)\), and \(T_c(t)\). These variables are related to the scale factor \(a_{nc}(t)\) and the anisotropy parameter \(\beta_{nc}(t)\) according to the transformations in Eqs. \eqref{EquivalencesNC}. Furthermore, the equations are highly coupled and non-linear, with the presence of terms involving \(\gamma\), which adjust the effects of anisotropy and corrections to the expansion of the universe. The constraint \(H_c = 0\) represents an additional condition that must be satisfied for the system to be consistent. 

\subsection{For the case of $\kappa =-1$ }
\label{sec:level 4.1} 

The system of equations \eqref{approx_dda}–\eqref{approx_Hc}, corresponding to the case \(\kappa = -1\),  which represents a BIII cosmological model with open spatial curvature, was numerically solved due to the intrinsic complexity of the expressions involved. To this end, the initial conditions \(a_c(0)\), \(\dot{a}_c(0)\), \(\beta_c(0)\), \(\dot{\beta}_c(0)\), and \(T_c(0)\) were specified, along with the values of the NC parameter \(\gamma\) and the integration constant \(C\). Then it was verified that these initial values satisfied the dynamical constraint \(H_c = 0\), as required by the Hamiltonian formalism. This constraint must hold for all times $t$, ensuring the consistency of the system’s evolution within the canonical framework.

In practice, the procedure was carried out by treating \(\dot{\beta}_c(0)\) as a free variable while fixing the other parameters, adjusting its value so that equation \eqref{approx_Hc} was fulfilled. Alternatively, if \(\dot{\beta}_c(0) \) was chosen instead as fixed, the constant \(C\) was determined accordingly to satisfy the constraint. The choice between treating \( C\) or \(\dot{\beta}_c(0) \) as the adjustable parameter is arbitrary and does not affect the validity or generality of the numerical solution obtained.

The numerical solutions for the variables \(a_c(t)\), \(\beta_c(t)\), and \(T_c(t)\) enable the reconstruction of the physical variables through the transformations defined in \eqref{EquivalencesNC}. These are the scale factor \(a_{nc}(t) = a_c + (\gamma/2) T_c\), which characterizes the expansion of the universe, and the anisotropy parameter \(\beta_{nc}(t) = \beta_c\), which encodes the anisotropic features present in the early stages of cosmic evolution. A detailed analysis of the behavior of these physical variables will be presented in the following sections. The graphs that we are going to present
are examples that show the general behavior of the solutions in a clear way. In these examples, we choose the numerical values of the parameters and initial conditions to show the different behaviors in the clearest way.

\subsubsection{Varying the NC parameter $\gamma$}
\begin{figure}[H]
    \centering
    \begin{subfigure}[b]{0.49\textwidth}
        \includegraphics[width=1\textwidth]{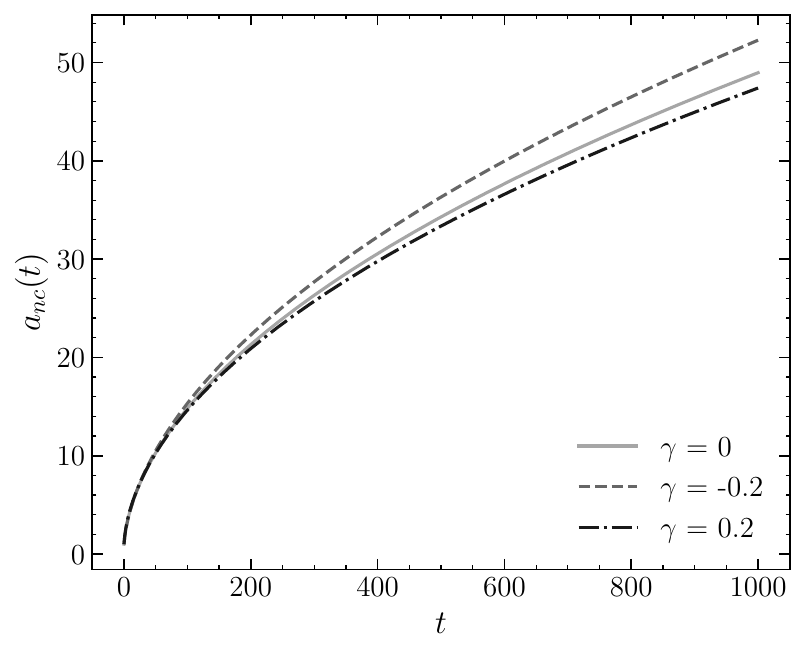} %
        \caption{ Evolution of the scale factor \(a_{nc}(t)\) as a function of time \(t\).} 
        \label{fig:BIII_a_gamma}
    \end{subfigure}
    \hfill
    \begin{subfigure}[b]{0.49\textwidth}
        \includegraphics[width=1\textwidth]{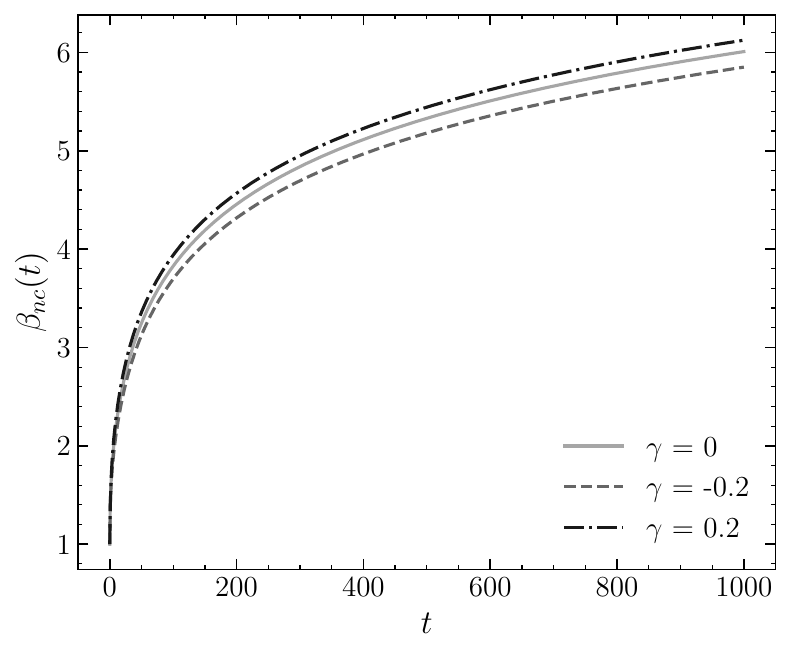} %
        \caption{Behavior of the anisotropy parameter \(\beta_{nc}(t)\) as a function of time \(t\).}
        \label{fig:BIII_beta_gamma}
    \end{subfigure}
    \caption{\justifying The graphs (a) and (b) show the numerical solutions obtained under initial conditions: \(a_c(0) = 1\), \(\dot{a}_c(0) = 1\), \(\beta_c(0) = 1\),  \(\dot{\beta}_c(0) = 1\), and \(T_c(0) = 0\). For each selected value of \(\gamma= \{0,\, -0.2,\, 0.2\}\), the corresponding value of \(C  = \{ 6.58747, \, 5.43337, \, 7.74158\}\) was calculated using the constraint \(H_c = 0\), while keeping the other initial conditions fixed.}
    \label{fig:BIII_varying_gamma}
\end{figure}

In \cref{fig:BIII_varying_gamma}, an analysis of the effect of the NC parameter \( \gamma \) on the evolution of the scale factor \( a_{nc}(t) \) and the anisotropy parameter \( \beta_{nc}(t) \) is initiated. In \cref{table:BIII_varying_gamma}, numerical data on its behavior at different time instants are presented, showing how the different values of \( \gamma \) influence the expansion of the universe and the stabilization of anisotropy in late times.

Specifically, with respect to the evolution of the scale factor \( a_{nc}(t) \), we observe from \cref{fig:BIII_a_gamma} that the solutions are always expansive. On the other hand, depending on the value of the NC parameter \( \gamma \), we identify a subtle but noticeable effect. From \cref{table:BIII_varying_gamma}, we observe that: for \( t = 100000 \), the scale factor reaches the value of \( 499.3741 \) when \( \gamma = 0 \), \( 553.1807 \) when \( \gamma = -0.2 \), and \( 473.9523 \) when \( \gamma = 0.2 \). These results indicate that a negative value of \( \gamma \) slightly increases the expansion, while a positive value of \( \gamma \) slightly reduces it. Therefore, the expansion rate is slightly more pronounced for negative values of \( \gamma \), suggesting that the parameter influences the rate of growth of the universe. This behavior can be attributed to the way the NC parameter affects the energy distribution within the model, potentially modifying the effective pressure of the system, which in turn influences the expansion rate.

As for the anisotropy parameter \( \beta_{nc}(t) \), we observe from \cref{fig:BIII_beta_gamma} that this parameter shows rapid growth in the initial stages, followed by stabilization toward a more stable value that varies with \( \gamma \). From \cref{table:BIII_varying_gamma}, we observe that: for \( t = 100000 \), the obtained values are \( 10.60 \) for \( \gamma = 0 \), \( 10.39 \) for \( \gamma = -0.2 \), and \( 10.73 \) for \( \gamma = 0.2 \). These results show that a positive value of \( \gamma \) slightly increases the residual anisotropy, while a negative value of \( \gamma \) reduces it. Furthermore, the rate of change \( \dot{\beta}_{nc}(t) \) decreases progressively over time, indicating that the anisotropy tends to a constant in the late-time regime. For \( t = 100000 \), the values of \( \dot{\beta}_{nc}(t) \) are \( 0.00001 \) for all cases, confirming that the evolution of anisotropy becomes practically negligible at late times, regardless of the value of \( \gamma \).

\begin{table}[H]
\centering
\begin{tabular}{lS[table-format=8.0e+7] S[table-format=6.0e+6] S[table-format=6.0e+6] S[table-format=8.0e+6]} 
\toprule
{\(t\)} & {\(a_{nc}(t)\)} & {\(\beta_{nc}(t)\)} & {\(\dot{\beta}_{nc}(t)\)} \\ 
\midrule
0 & 1 & 1 & 1 \\
6765 & 129.164429 & 7.90581 & 0.000147 \\
10946 & 164.586699 & 8.386061 & 0.000091 \\
17711 & 209.632437 & 8.866669 & 0.000056 \\
28657 & 266.918407 & 9.347501 & 0.000035 \\
46368 & 339.773622 & 9.828475 & 0.000022 \\
75025 & 432.432997 & 10.309537 & 0.000013 \\
100000 & 499.374086 & 10.596825 & 0.000010 \\
\midrule 
0 & 1 & 1 & 1 \\
6765 & 140.504186 & 7.724368 & 0.000146 \\
10946 & 179.714842 & 8.199524 & 0.00009 \\
17711 & 229.709761 & 8.675289 & 0.000056 \\
28657 & 293.446562 & 9.151532 & 0.000035 \\
46368 & 374.692907 & 9.628163 & 0.000021 \\
75025 & 478.247856 & 10.10512 & 0.000013 \\
100000 & 553.180714 & 10.390063 & 0.000010 \\
\midrule
0 & 1 & 1 & 1 \\
6765 & 123.850819 & 8.030033 & 0.000148 \\
10946 & 157.475065 & 8.51284 & 0.000092 \\
17711 & 200.172186 & 8.995896 & 0.000057 \\
28657 & 254.398652 & 9.479035 & 0.000035 \\
46368 & 323.278445 & 9.962151 & 0.000022 \\
75025 & 410.784542 & 10.445179 & 0.000013 \\
100000 & 473.952318 & 10.733554 & 0.000010 \\
\bottomrule
\end{tabular}
\caption{Results of the physical variables shown in \cref{fig:BIII_varying_gamma}, for different time instances \(t\), with \(\gamma = 0\) in the upper part, \(\gamma = -0.2\) in the central part,  and \(\gamma = 0.2\) in the lower part.}
\label{table:BIII_varying_gamma}
\end{table}

\subsubsection{Varying the energy density $C$ }

\begin{figure}[H]
    \centering
    \begin{subfigure}[b]{0.49\textwidth}
        \includegraphics[width=1\textwidth]{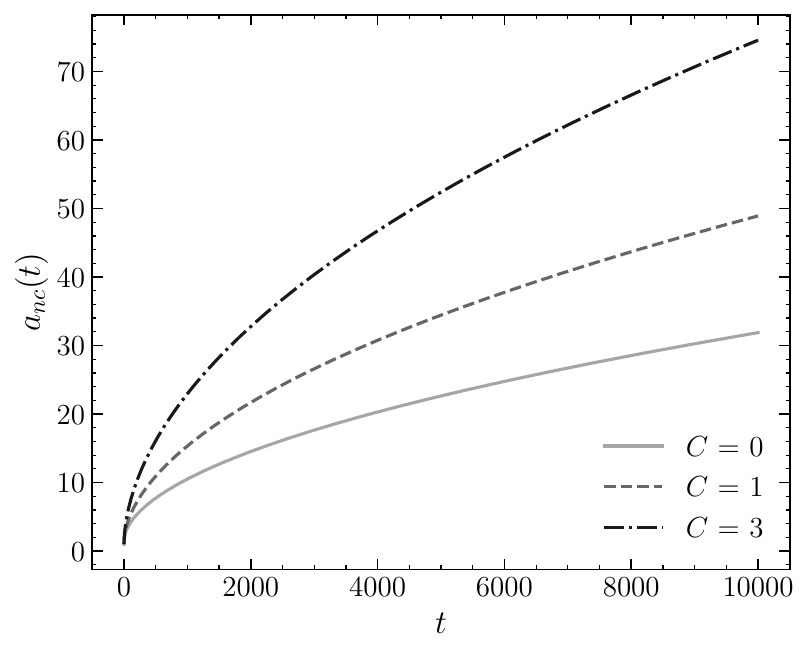} %
        \caption{ Evolution of the scale factor \(a_{nc}(t)\) as a function of time \(t\).} 
        \label{fig:BIII_a_C}
    \end{subfigure}
    \hfill
    \begin{subfigure}[b]{0.49\textwidth}
        \includegraphics[width=1\textwidth]{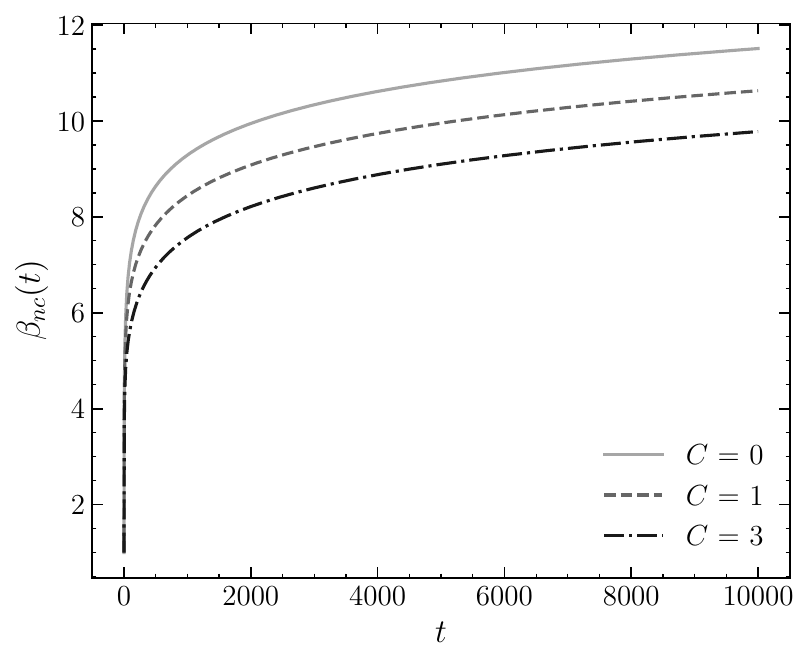} %
        \caption{Behavior of the anisotropy parameter \(\beta_{nc}(t)\) as a function of time \(t\).}
        \label{fig:BIII_beta_C}
    \end{subfigure}
    \caption{\justifying The graphs (a) and (b) show the numerical solutions obtained under initial conditions: \(a_c(0) = 1\), \(\dot{a}_c(0) = 1\), \(\beta_c(0) = 1\), \(T_c(0) = 0\), and the NC parameter \(\gamma = -0.01\). For each selected value of \(C= \{0,\, 1,\, 3\}\), the corresponding value of \(\dot{\beta}_c(0)  = \{ 5.78536, \, 5.50429, \, 4.8607\}\) was calculated using the constraint \(H_c = 0\), while keeping the other initial conditions fixed.}
    \label{fig:BIII_varying_C}
\end{figure}

Next, the effect of varying the energy density \( C \) is studied. As shown in \cref{fig:BIII_a_C}, an increase in \( C \) leads to a more accelerated growth of the scale factor \( a_{nc}(t) \), indicating that a higher energy density in the system significantly influences the rate of expansion of the universe. This effect becomes particularly evident in later times, where the curves clearly separate and highlight a strong dependence of expansion on the value of \( C \). Regarding the evolution of anisotropy \( \beta_{nc}(t) \), we observe from \cref{fig:BIII_beta_C} that this parameter shows rapid growth in the early stages, followed by a smooth stabilization toward an asymptotic regime. Furthermore, as \( C \) increases, the value of anisotropy decreases.

\begin{table}[H]
\centering
\begin{tabular}{lS[table-format=8.0e+7] S[table-format=6.0e+6] S[table-format=6.0e+6] S[table-format=8.0e+6]} 
\toprule
{\(t\)} & {\(a_{nc}(t)\)} & {\(\beta_{nc}(t)\)} & {\(\dot{\beta}_{nc}(t)\)} \\ 
\midrule
0 & 1 & 1 & 5.785361 \\
6765 & 26.27245 & 11.128413 & 0.000144 \\
10946 & 33.34535 & 11.599695 & 0.00009 \\
17711 & 42.381682 & 12.073095 & 0.000056 \\
28657 & 53.915827 & 12.548111 & 0.000034 \\
46368 & 68.627935 & 13.024344 & 0.000021 \\
75025 & 87.383781 & 13.501489 & 0.000013 \\
100000 & 100.955147 & 13.786749 & 0.00001 \\
\midrule 
0 & 1 & 1 & 4.860701 \\
6765 & 61.116958 & 9.394866 & 0.000145 \\
10946 & 78.064725 & 9.868631 & 0.00009 \\
17711 & 99.641000 & 10.344768 & 0.000056 \\
28657 & 127.100693 & 10.822483 & 0.000035 \\
46368 & 162.041345 & 11.301244 & 0.000021 \\
75025 & 206.496301 & 11.7807 & 0.000013 \\
100000 & 238.619806 & 12.067232 & 0.00001 \\
\bottomrule
\end{tabular}
\caption{Results of the physical variables shown in \cref{fig:BIII_varying_C}, for different time instances \(t\), with \(C = 0\) in the upper part and \(C = 3\) in the lower part.}
\label{table:BIII_varying_C}
\end{table}

The numerical analysis presented in \cref{table:BIII_varying_C}, shows that for \( t = 100000 \), the scale factor reaches a value of \( 100.9551 \) when \( C = 0 \), and \( 238.6198 \) when \( C = 3 \), confirming that higher values of \( C \) significantly intensify the expansion of \( a_{nc}(t) \). On the other hand, the anisotropy parameter \( \beta_{nc}(t) \) tends to stabilize at finite values of \( 13.79 \) for \( C = 0 \) and \( 12.07 \) for \( C = 3 \), indicating that the anisotropy does not disappear, but its magnitude reduces as the energy density increases. Furthermore, the rate of change \( \dot{\beta}_{nc}(t) \) decreases rapidly over time, reaching values on the order of \( 1.0 \times 10^{-5} \) at \( t = 100000 \) in both cases. This suggests that anisotropy evolves toward an equilibrium state as time progresses, although the value at which it stabilizes depends on the value of \( C \).

\subsubsection{Varying the  \(a_{c}(0)\) }

\begin{figure}[H]
    \centering
    \begin{subfigure}[b]{0.49\textwidth}
        \includegraphics[width=1\textwidth]{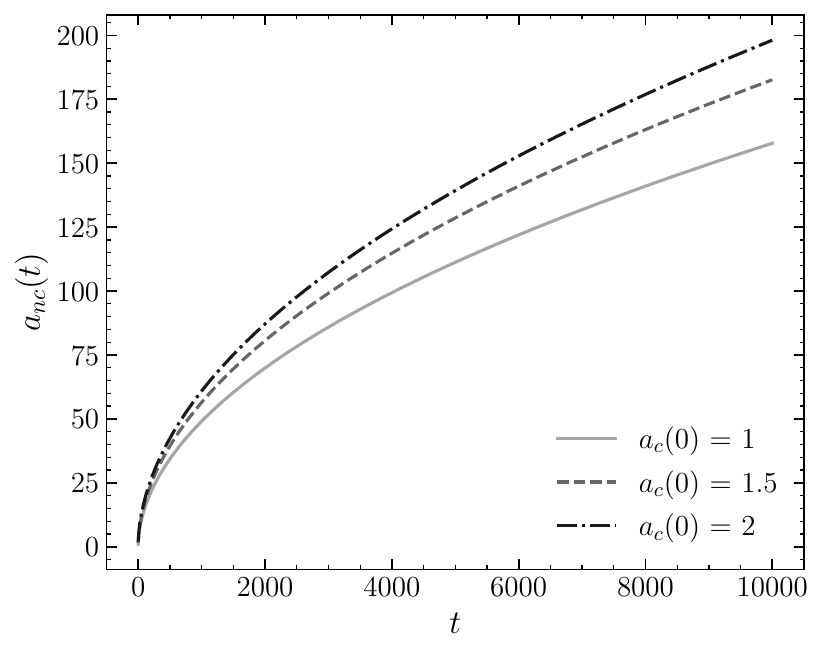} %
        \caption{ Evolution of the scale factor \(a_{nc}(t)\) as a function of time \(t\).} 
        \label{fig:BIII_a_a.pdf}
    \end{subfigure}
    \hfill
    \begin{subfigure}[b]{0.49\textwidth}
        \includegraphics[width=1\textwidth]{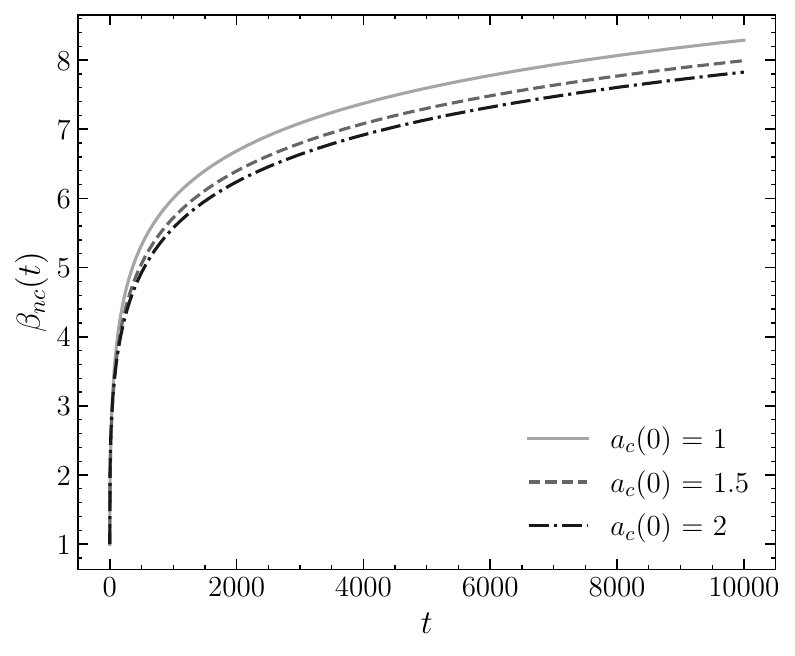} %
        \caption{Behavior of the anisotropy parameter \(\beta_{nc}(t)\) as a function of time \(t\).}
        \label{fig:BIII_beta_a.pdf}
    \end{subfigure}
    \caption{\justifying The graphs (a) and (b) show the numerical solutions obtained under initial conditions: \(\dot{a}_c(0) = 1\), \(\beta_c(0) = 1\), \(\dot{\beta}_c(0) = 1\), \(T_c(0) = 0\), and the NC parameter \(\gamma = -0.01\). For each selected value of \(a_c(0) = \{1,\, 1.5,\, 2\}\), the corresponding value of \(C = \{6.5298, \, 15.5508, \, 28.1657\}\) was calculated using the constraint \(H_c = 0\), while keeping the other initial conditions fixed.}
    \label{fig:BIII_varying_a}
\end{figure}

The effect of varying the initial condition \( a_c(0) \) on the evolution of the scale factor \( a_{nc}(t) \) and the anisotropy parameter \( \beta_{nc}(t) \) is examined next. As shown in \cref{fig:BIII_varying_a}, the influence of \( a_c(0) \) on both variables is illustrated for the values \( a_c(0) = 1 \), \( 1.5 \), and \( 2 \). The results show that an increase in \( a_c(0) \) leads to greater growth of the scale factor, implying a more accelerated expansion of the universe. Regarding the anisotropy parameter \( \beta_{nc}(t) \), its evolution shows rapid growth in the early stages, followed by a progressive stabilization toward finite values. Interestingly, as \( a_c(0) \) increases, the asymptotic value of \( \beta_{nc}(t) \) decreases slightly.

The numerical analysis of these results, presented in \cref{table:BIII_varying_a}, shows that for \( t = 100000 \), the scale factor reaches a value of \( 501.4288 \) when \( a_c(0) = 1 \), and \( 632.0084 \) when \( a_c(0) = 2 \), confirming that the expansion rate increases with higher initial values of \( a_c(0) \). On the other hand, the anisotropy parameter \( \beta_{nc}(t) \) follows a progressive stabilization at later times. In all cases, anisotropy persists at finite values, although its final value depends sensitively on the initial value of \( a_c(0) \). For \( t = 100000 \), it tends to values of \( 10.6 \) for \( a_c(0) = 1 \), and \( 10.1 \) for \( a_c(0) = 2 \), indicating that anisotropy persists at finite levels rather than disappearing. Furthermore, the rate of change \( \dot{\beta}_{nc}(t) \) decreases significantly over time in both cases, reaching values on the order of \( 1.0 \times 10^{-5} \) for \( t = 100000 \), suggesting that anisotropy evolves toward an equilibrium state where its variation becomes practically negligible.

\begin{table}[H]
\centering
\begin{tabular}{lS[table-format=8.0e+7] S[table-format=6.0e+6] S[table-format=6.0e+6] S[table-format=8.0e+6]}  
\toprule
{\(t\)} & {\(a_{nc}(t)\)} & {\(\beta_{nc}(t)\)} & {\(\dot{\beta}_{nc}(t)\)} \\ 
\midrule
0 & 1 & 1 & 1 \\
6765 & 129.593091 & 7.89773 & 0.000147 \\
10946 & 165.160251 & 8.377767 & 0.000091 \\
17711 & 210.395436 & 8.858171 & 0.000056 \\
28657 & 267.928457 & 9.338811 & 0.000035 \\
46368 & 341.105088 & 9.819604 & 0.000022 \\
75025 & 434.181749 & 10.300497 & 0.000013 \\
100000 & 501.428814 & 10.58769 & 0.000010 \\
\midrule 
0 & 2 & 1 & 1 \\
6765 & 162.470187 & 7.438364 & 0.000147 \\
10946 & 207.389223 & 7.916107 & 0.000091 \\
17711 & 264.520029 & 8.395069 & 0.000056 \\
28657 & 337.178437 & 8.874812 & 0.000035 \\
46368 & 429.584201 & 9.355054 & 0.000022 \\
75025 & 547.106591 & 9.835616 & 0.000013 \\
100000 & 632.008434 & 10.122679 & 0.000010 \\
\bottomrule
\end{tabular}
\caption{Results of the physical variables shown in \cref{fig:BIII_varying_a}, for different time instances \(t\), with \(a_c(0) = 1\) in the upper part and \(a_c(0) = 2\) in the lower part.}
\label{table:BIII_varying_a}
\end{table}

\subsubsection{Varying the \(\dot{a}_{c}(0)\) }

\begin{figure}[H]
    \centering
    \begin{subfigure}[b]{0.49\textwidth}
        \includegraphics[width=1\textwidth]{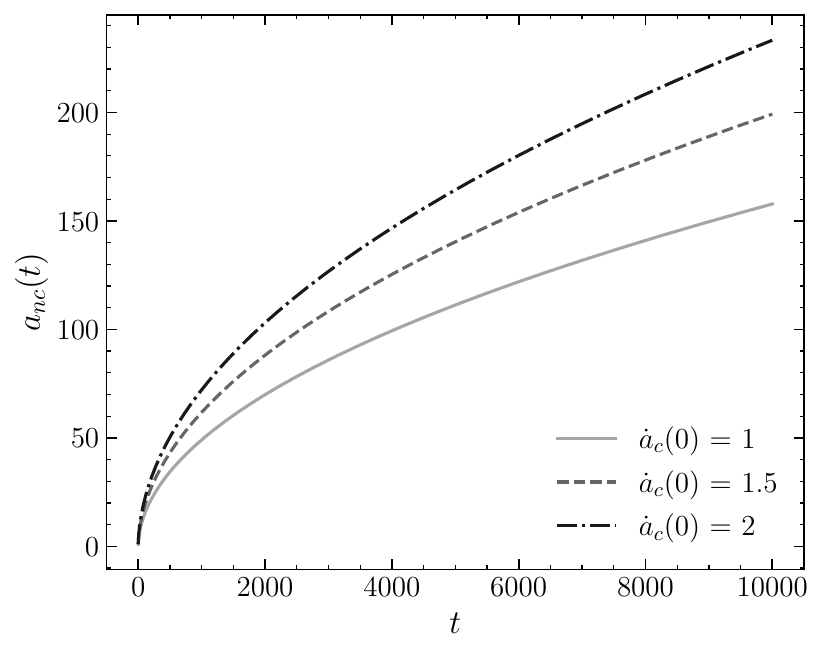} %
        \caption{ Evolution of the scale factor \(a_{nc}(t)\) as a function of time \(t\).} 
        \label{fig:BIII_a_da.pdf}
    \end{subfigure}
    \hfill
    \begin{subfigure}[b]{0.49\textwidth}
        \includegraphics[width=1\textwidth]{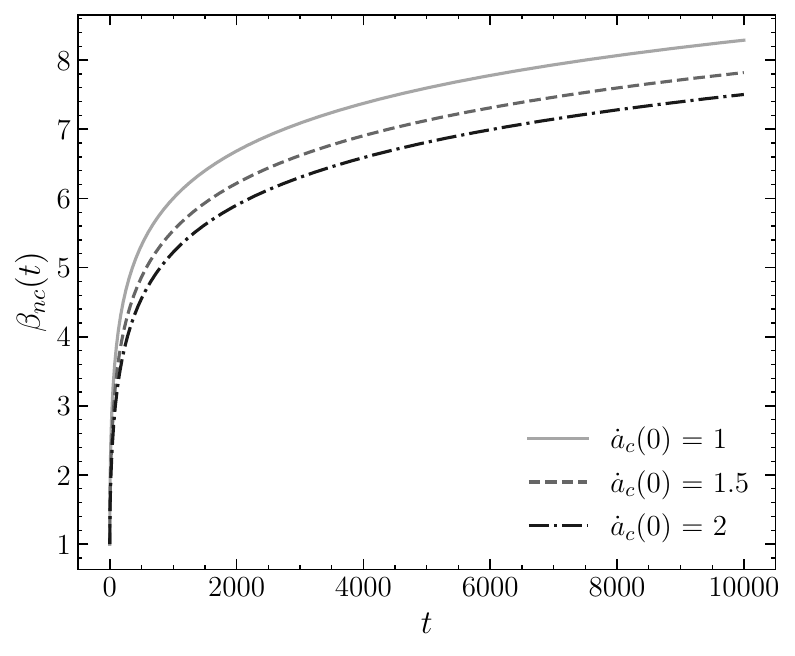} %
        \caption{Behavior of the anisotropy parameter \(\beta_{nc}(t)\) as a function of time \(t\).}
        \label{fig:BIII_beta_da.pdf}
    \end{subfigure}
    \caption{\justifying The graphs (a) and (b) show the numerical solutions obtained under initial conditions: \(a_c(0) = 1\), \(\beta_c(0) = 1\), \(\dot{\beta}_c(0) = 1\), \(T_c(0) = 0\), and the NC parameter \(\gamma = -0.01\). For each selected value of \(\dot{a}_c(0) = \{1,\, 1.5,\, 2\}\), the corresponding value of \(C = \{ 6.5298, \, 14.7829, \,  25.9576\}\) was calculated using the constraint \(H_c = 0\), while keeping the other initial conditions fixed.}
    \label{fig:BIII_varying_da}
\end{figure}

\begin{table}[H]
\centering
\begin{tabular}{lS[table-format=8.0e+7] S[table-format=6.0e+6] S[table-format=6.0e+6] S[table-format=8.0e+6]}
\toprule
{\(t\)} & {\(a_{nc}(t)\)} & {\(\beta_{nc}(t)\)} & {\(\dot{\beta}_{nc}(t)\)} \\ 
\midrule
0 & 1 & 1 & 1 \\
6765 & 129.593091 & 7.89773 & 0.000147 \\
10946 & 165.160251 & 8.377767 & 0.000091 \\
17711 & 210.395436 & 8.858171 & 0.000056 \\
28657 & 267.928457 & 9.338811 & 0.000035 \\
46368 & 341.105088 & 9.819604 & 0.000022 \\
75025 & 434.181749 & 10.300497 & 0.000013 \\
100000 & 501.428814 & 10.58769 & 0.000010 \\
\midrule 
0 & 1 & 1 & 1 \\
6765 & 191.503249 & 7.113079 & 0.000147 \\
10946 & 244.210567 & 7.592824 & 0.000091 \\
17711 & 311.234571 & 8.073087 & 0.000056 \\
28657 & 396.468117 & 8.553675 & 0.000035 \\
46368 & 504.864111 & 9.034466 & 0.000022 \\
75025 & 642.723662 & 9.515387 & 0.000013 \\
100000 & 742.319000 & 9.802604 & 0.000010 \\
\bottomrule
\end{tabular}
\caption{Results of the physical variables shown in \cref{fig:BIII_varying_da}, for different time instances \(t\), with \(\dot{a}_c(0) = 1\) in the upper part and \(\dot{a}_c(0) = 2\) in the lower part.}
\label{table:BIII_varying_da}
\end{table}

Continuing with the analysis of the initial conditions, the effect of varying the initial condition \( \dot{a}_c(0) \) on the evolution of the system is examined. As shown in \cref{fig:BIII_varying_da}, the results indicate that a higher initial value of \( \dot{a}_c(0) \) leads to a faster growth of the scale factor \( a_{nc}(t) \), promoting a more accelerated expansion of the universe. In all cases, the evolution of \( a_{nc}(t) \) presents a steeper slope as the value of \( \dot{a}_c(0) \) increases. Regarding the behavior of the anisotropy parameter \( \beta_{nc}(t) \), its evolution begins with a sharp increase in the initial stages, followed by a stabilization toward finite values. Furthermore, as \( \dot{a}_c(0) \) increases, this asymptotic value decreases slightly, suggesting that higher initial expansion velocities could favor a more efficient isotropization at later times.

The numerical results, presented in \cref{table:BIII_varying_da}, show that for \( t = 100000 \), the scale factor reaches a value of \( 501.4288 \) when \( \dot{a}_c(0) = 1 \), and \( 742.3190 \) when \( \dot{a}_c(0) = 2 \), confirming that a higher initial value of \( \dot{a}_c(0) \) increases the expansion rate of \( a_{nc}(t) \). On the other hand, the anisotropy \( \beta_{nc}(t) \) follows a progressive stabilization at later times. In all cases, anisotropy persists at finite values, although its final value depends sensitively on the initial value of \( \dot{a}_c(0) \). For \( t = 100000 \), it tends to values of \( 10.6 \) for \( \dot{a}_c(0) = 1 \), and \( 9.8 \) for \( \dot{a}_c(0) = 2 \), indicating a reduction in anisotropy and its stabilization at finite values. Similarly, the rate of change \( \dot{\beta}_{nc}(t) \) decreases rapidly with time, reaching values on the order of \( 1.0 \times 10^{-5} \) for both cases at later times, suggesting that anisotropy evolves toward an equilibrium state where its variation becomes negligible.

\subsubsection{Varying the  \(\beta_{c}(0)\)}

\begin{figure}[H]
    \centering
    \begin{subfigure}[b]{0.49\textwidth}
        \includegraphics[width=1\textwidth]{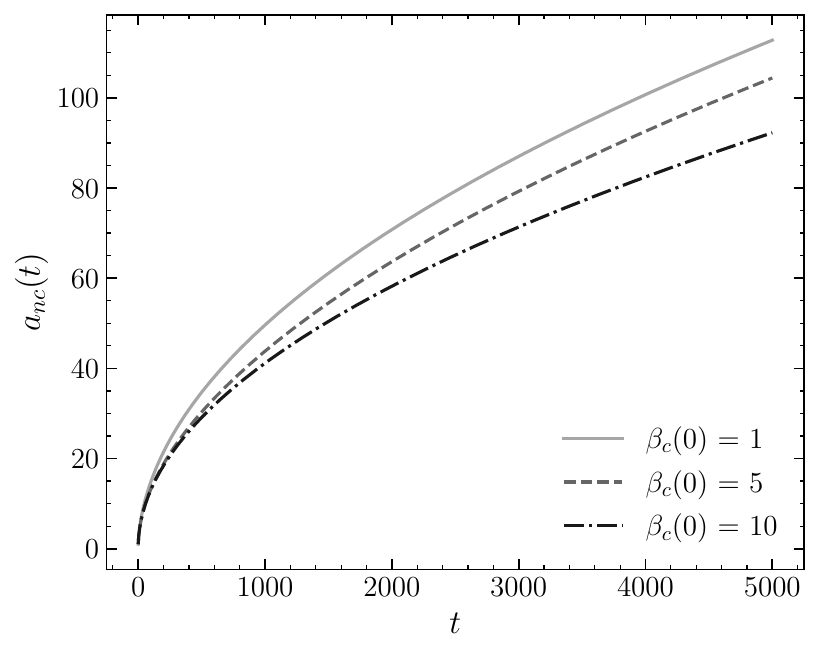} %
        \caption{ Evolution of the scale factor \(a_{nc}(t)\) as a function of time \(t\).} 
        \label{fig:BIII_a_b.pdf}
    \end{subfigure}
    \hfill
    \begin{subfigure}[b]{0.49\textwidth}
        \includegraphics[width=1\textwidth]{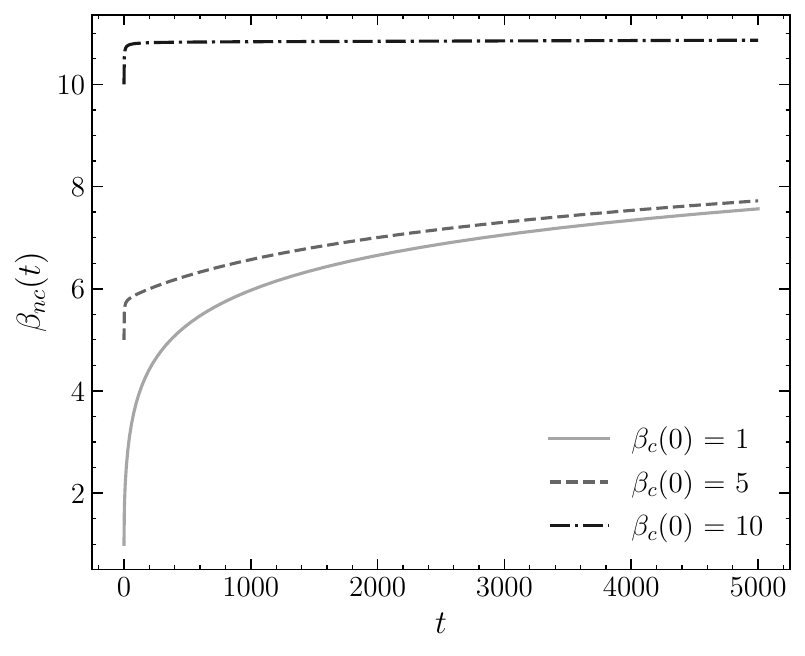} %
        \caption{Behavior of the anisotropy parameter \(\beta_{nc}(t)\) as a function of time \(t\).}
        \label{fig:BIII_beta_b.pdf}
    \end{subfigure}
    \caption{\justifying The graphs (a) and (b) show the numerical solutions obtained under initial conditions: \(a_c(0) = 1\), \(\dot{a}_c(0) = 1\), \(\dot{\beta}_c(0) = 1\), \(T_c(0) = 0\), and the NC parameter \(\gamma = -0.05\). For each selected value of \(\beta_c(0) = \{1,\, 5,\, 10\}\), the corresponding value of \(C = \{6.2990, \, 102.7980, \, 2920.6400\}\) was calculated using the constraint \(H_c = 0\), while keeping the other initial conditions fixed.}
    \label{fig:BIII_varying_b}
\end{figure}

\begin{table}[H]
\centering
\begin{tabular}{lS[table-format=8.0e+7] S[table-format=6.0e+6] S[table-format=6.0e+6] S[table-format=8.0e+6]}
\toprule
{\(t\)} & {\(a_{nc}(t)\)} & {\(\beta_{nc}(t)\)} & {\(\dot{\beta}_{nc}(t)\)} \\ 
\midrule
0 & 1 & 1 & 1 \\
6765 & 131.455764 & 7.864127 & 0.000147 \\
10946 & 167.650833 & 8.34325 & 0.000091 \\
17711 & 213.70686 & 8.822787 & 0.000056 \\
28657 & 272.310198 & 9.302609 & 0.000035 \\
46368 & 346.87927 & 9.782633 & 0.000022 \\
75025 & 441.763773 & 10.262806 & 0.000013 \\
100000 & 510.336552 & 10.54959 & 0.00001 \\
\midrule 
0 & 1 & 10 & 1 \\
6765 & 107.443484 & 10.877028 & 0.000007 \\
10946 & 137.053759 & 10.905822 & 0.000007 \\
17711 & 175.07236 & 10.951388 & 0.000007 \\
28657 & 224.079603 & 11.021804 & 0.000006 \\
46368 & 287.549835 & 11.127043 & 0.000006 \\
75025 & 370.138952 & 11.277662 & 0.000005 \\
100000 & 431.032171 & 11.39297 & 0.000004 \\
\bottomrule
\end{tabular}
\caption{Results of the physical variables shown in \cref{fig:BIII_varying_b}, for different time instances \(t\), with \(\beta_c(0) = 1\) in the upper part and \(\beta_c(0) = 10\) in the lower part.}
\label{table:BIII_varying_b}
\end{table}

Next, the influence of the initial condition \( \beta_c(0) \) on the system's dynamics is evaluated. The results, shown in \cref{fig:BIII_varying_b}, reveal that an increase in \( \beta_c(0) \) leads to a slower growth of the scale factor \( a_{nc}(t) \), suggesting that a higher initial anisotropy acts as a source of geometric resistance to the expansion of the universe. On the other hand, the anisotropy \( \beta_{nc}(t) \) shows rapid growth in the early stages, followed by a smooth transition to an asymptotic regime. Furthermore, higher values of \( \beta_c(0) \) lead to a more pronounced residual anisotropy.

The numerical analysis of these results, presented in \cref{table:BIII_varying_b}, confirms these observations. For \( t = 100000 \), the scale factor reaches a value of \( 510.3366 \) when \( \beta_c(0) = 1 \), but only \( 431.0322 \) when \( \beta_c(0) = 10 \), showing a decrease, highlighting the inhibiting effect of the initial anisotropy on the expansion. On the other hand, \( \beta_{nc}(t) \) stabilizes approximately at values of \( 10.55 \) for \( \beta_c(0) = 1 \) and \( 11.39 \) for \( \beta_c(0) = 10 \), indicating not only that anisotropy does not disappear, but its final value directly depends on the initial conditions. Furthermore, the rate of change \( \dot{\beta}_{nc}(t) \) decreases significantly over time, reaching values on the order of \( 1.0 \times 10^{-5} \) and \( 4.0 \times 10^{-6} \), respectively, for the same values of \( \beta_c(0) \). This means that the system evolves toward an equilibrium isotropic state, where the directional deformation becomes dynamically negligible.

\subsubsection{Varying the  \(\dot{\beta}_{c}(0)\)}

\begin{figure}[H]
    \centering
    \begin{subfigure}[b]{0.49\textwidth}
        \includegraphics[width=1\textwidth]{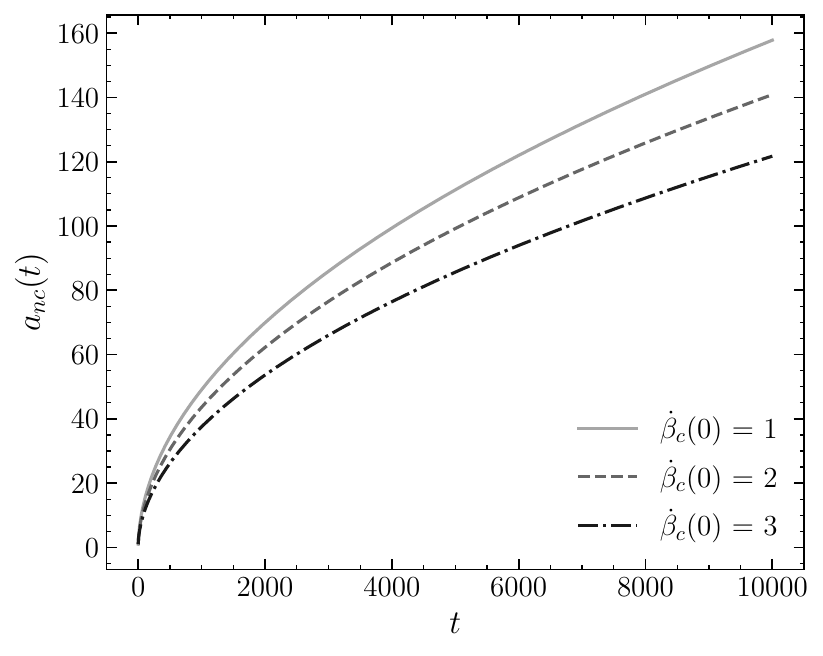} %
        \caption{ Evolution of the scale factor \(a_{nc}(t)\) as a function of time \(t\).} 
        \label{fig:BIII_a_db.pdf}
    \end{subfigure}
    \hfill
    \begin{subfigure}[b]{0.49\textwidth}
        \includegraphics[width=1\textwidth]{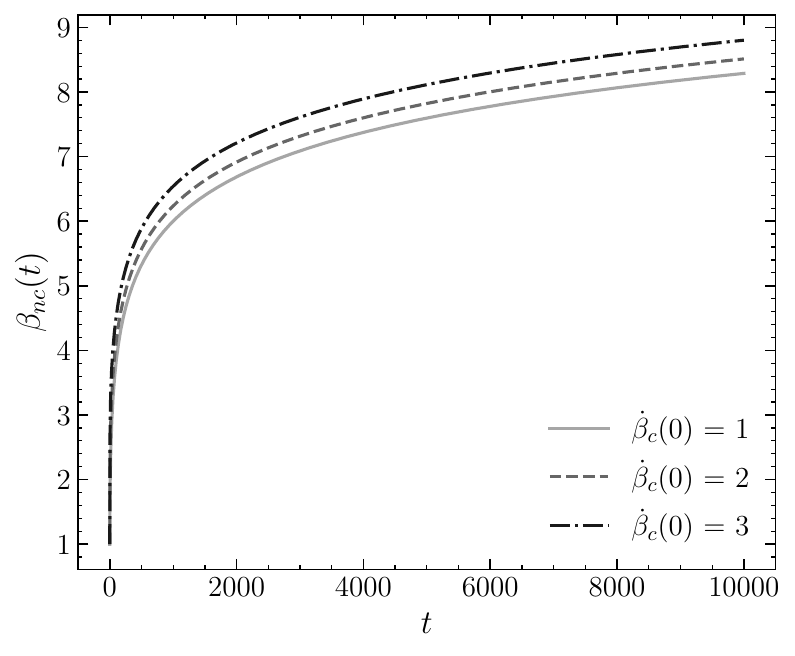} %
        \caption{Behavior of the anisotropy parameter \(\beta_{nc}(t)\) as a function of time \(t\).}
        \label{fig:BIII_beta_db.pdf}
    \end{subfigure}
    \caption{\justifying The graphs (a) and (b) show the numerical solutions obtained under initial conditions: \(a_c(0) = 1\), \(\dot{a}_c(0) = 1\), \(\beta_c(0) = 1\), \(T_c(0) = 0\), and the NC parameter \(\gamma = -0.01\). For each selected value of \(\dot{\beta}_c(0) = \{1,\, 2,\, 3\}\), the corresponding value of \(C = \{6.5298, \,7.0085, \,6.51328\}\) was calculated using the constraint \(H_c = 0\), while keeping the other initial conditions fixed.}
    \label{fig:BIII_varying_db}
\end{figure}

\begin{table}[H]
\centering
\begin{tabular}{lS[table-format=8.0e+7] S[table-format=6.0e+6] S[table-format=6.0e+6] S[table-format=8.0e+6]} 
\toprule
{\(t\)} & {\(a_{nc}(t)\)} & {\(\beta_{nc}(t)\)} & {\(\dot{\beta}_{nc}(t)\)} \\ 
\midrule
0 & 1 & 1 & 1 \\
6765 & 129.593091 & 7.89773 & 0.000147 \\
10946 & 165.160251 & 8.377767 & 0.000091 \\
17711 & 210.395436 & 8.858171 & 0.000056 \\
28657 & 267.928457 & 9.338811 & 0.000035 \\
46368 & 341.105088 & 9.819604 & 0.000022 \\
75025 & 434.181749 & 10.300497 & 0.000013 \\
100000 & 501.428814 & 10.58769 & 0.00001 \\
\midrule 
0 & 1 & 1 & 3 \\
6765 & 99.858193 & 8.41316 & 0.000147 \\
10946 & 127.423054 & 8.891245 & 0.000091 \\
17711 & 162.485774 & 9.370362 & 0.000056 \\
28657 & 207.083575 & 9.850146 & 0.000035 \\
46368 & 263.808942 & 10.330363 & 0.000022 \\
75025 & 335.96054 & 10.810861 & 0.000013 \\
100000 & 388.089198 & 11.097877 & 0.00001 \\
\bottomrule
\end{tabular}
\caption{Results of the physical variables shown in \cref{fig:BIII_varying_db}, for different time instances \(t\), with \(\dot{\beta}_c(0) = 1\) in the upper part and \(\dot{\beta}_c(0) = 3\) in the lower part.}
\label{table:BIII_varying_db}
\end{table}

Finally, the role of the initial anisotropy rate of change \( \dot{\beta}_c(0) \) is analyzed. As shown in \cref{fig:BIII_varying_db}, a higher value of \( \dot{\beta}_c(0) \) slows down the growth of the scale factor \( a_{nc}(t) \), although it does not alter its overall expansive nature. This behavior suggests that a stronger initial anisotropic variation acts as an effective resistance to the expansion of the universe. Regarding the evolution of the anisotropy parameter \( \beta_{nc}(t) \), a sharp increase is observed in the early stages, followed by a smooth stabilization toward an asymptotic regime. Furthermore, as \( \dot{\beta}_c(0) \) increases, this asymptotic value also increases.

From the numerical results presented in \cref{table:BIII_varying_db}, it is observed that for \( t = 100000 \), the scale factor reaches a value of \( 501.4288 \) when \( \dot{\beta}_c(0) = 1 \), and \( 388.0892 \) when \( \dot{\beta}_c(0) = 3 \). This confirms that higher values of \( \dot{\beta}_c(0) \) reduce the expansion rate of \( a_{nc}(t) \). On the other hand, the anisotropy \( \beta_{nc}(t) \) tends to stabilize at approximately \( 10.59 \) for \( \dot{\beta}_c(0) = 1.0 \) and \( 11.10 \) for \( \dot{\beta}_c(0) = 3 \), reinforcing the idea that the anisotropy does not disappear but rather stabilizes. Furthermore, the rate of change \( \dot{\beta}_{nc}(t) \) decreases significantly over time, reaching values on the order of \( 1.0 \times 10^{-5} \) in both cases. This confirms that anisotropy evolves toward an equilibrium state in late times.


\subsection{For the case of $\kappa = 0$}
\label{sec:level 4.2} 

In this case, the system of equations \eqref{approx_dda}-\eqref{approx_Hc} was solved for a BI cosmological model with no curvature, specifically \(\kappa = 0\), using the same approach as in the previous subsection. The initial conditions \(a_c(0)\), \(\dot{a}_c(0)\), \(\beta_c(0)\), \(\dot{\beta}_c(0)\), and \(T_c(0)\) were defined, together with the values of the parameters \(C\) and \(\gamma\), to ensure that the restriction due to equation \eqref{approx_Hc} was satisfied. Based on the numerical solutions obtained for the commutative variables \(a_c(t)\), \(\beta_c(t)\), and \(T_c(t)\), the physical variables related to the expansion of the universe, \(a_{nc}(t) = a_c + \frac{\gamma}{2} T_c\), and the anisotropy parameter, \(\beta_{nc}(t) = \beta_c\), were calculated. Then, a detailed analysis of these physical variables was carried out, including their asymptotic behavior and consistency with an isotropic radiation-dominated universe. The graphs that we are going to present are examples that show the general behavior of the solutions in a clear way. In these examples, we choose the numerical values of the parameters and initial conditions to show the different behaviors in the clearest way.

\subsubsection{Varying the NC parameter $\gamma$}

\begin{figure}[H]
    \centering
    \begin{subfigure}[b]{0.49\textwidth}
        \includegraphics[width=1\textwidth]{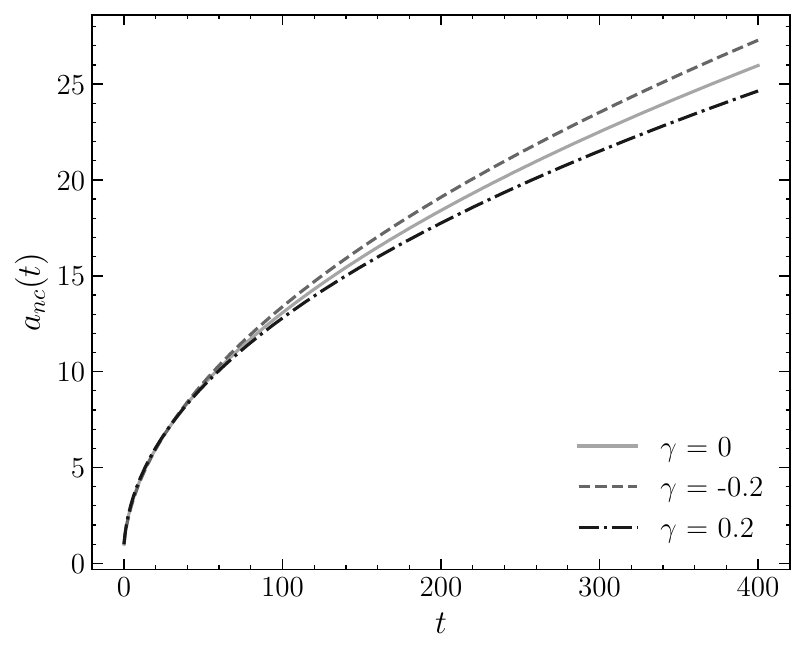} %
        \caption{ Evolution of the scale factor \(a_{nc}(t)\) as a function of time \(t\).} 
        \label{fig:BI_a_gamma}
    \end{subfigure}
    \hfill
    \begin{subfigure}[b]{0.49\textwidth}
        \includegraphics[width=1\textwidth]{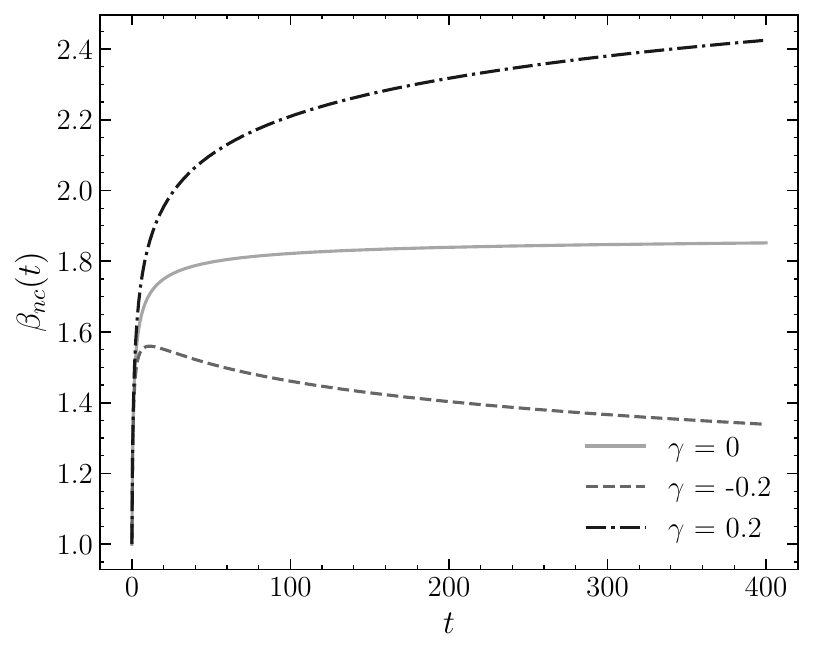} %
        \caption{Behavior of the anisotropy parameter \(\beta_{nc}(t)\) as a function of time \(t\).}
        \label{fig:BI_beta_gamma}
    \end{subfigure}
    \caption{\justifying The graphs (a) and (b) show the numerical solutions obtained under initial conditions: \(a_c(0) = 1\), \(\dot{a}_c(0) = 1\), \(\beta_c(0) = 1\),  \(\dot{\beta}_c(0) = 1\), and \(T_c(0) = 0\). For each selected value of \(\gamma= \{0,\, -0.2,\, 0.2\}\), the corresponding value of \(C  = \{7.3040, \, 6.1499, \, 8.4581\}\) was calculated using the constraint \(H_c = 0\), while keeping the other initial conditions fixed.}
    \label{fig:BI_varying_gamma}
\end{figure}

To begin the analysis of the flat model \((\kappa = 0)\), the impact of the NC parameter \( \gamma \) on the evolution of the scale factor \( a_{nc}(t) \) and the anisotropy parameter \( \beta_{nc}(t) \) is studied, as shown in \cref{fig:BI_varying_gamma}, along with the numerical results presented in \cref{table:BI_varying_gamma}. These results illustrate how different values of \( \gamma \) influence both the expansion rate of the universe and the stabilization of anisotropies at later times.

In particular, regarding the evolution of the scale factor \( a_{nc}(t) \), it is observed that the presence of the NC parameter \( \gamma \) modifies the expansion dynamics of the universe. Although all cases show an accelerated growth, the expansion is more pronounced for \( \gamma = -0.2 \), while it is slightly smaller for \( \gamma = 0.2 \), compared to the commutative case \( \gamma = 0 \). From \cref{table:BI_varying_gamma}, we observe that for \( t = 100000 \), the scale factor reaches a value of \( 408.3784 \) when \( \gamma = 0 \), \( 475.2813 \) for \( \gamma = -0.2 \), and \( 334.6021 \) for \( \gamma = 0.2 \). These results indicate that negative values of \( \gamma \) increase the expansion rate, while positive values tend to reduce it, confirming that the \( \gamma \) parameter has a significant effect on the evolution of \( a_{nc}(t) \). The parameter \( \gamma \) encapsulates geometric deformations that modify the dynamic equations of the early universe, affecting the expansion rate even in the absence of additional exotic components.

\begin{table}[H]
\centering
\begin{tabular}{lS[table-format=8.0e+7] S[table-format=6.0e+6] S[table-format=6.0e+6] S[table-format=8.0e+6]}
\toprule
{\(t\)} & {\(a_{nc}(t)\)} & {\(\beta_{nc}(t)\)} & {\(\dot{\beta}_{nc}(t)\)} \\ 
\midrule
0 & 1 & 1 & 1 \\
6765 & 106.317581 & 1.874408 & 5.374146e-07 \\
10946 & 135.200635 & 1.875963 & 2.611264e-07 \\
17711 & 171.940933 & 1.877186 & 1.268772e-07 \\
28657 & 218.675703 & 1.878147 & 6.164690e-08 \\
46368 & 278.123559 & 1.878902 & 2.995266e-08 \\
75025 & 353.742648 & 1.879496 & 1.455317e-08 \\
100000 & 408.37842 & 1.879789 & 9.457318e-09 \\
\midrule 
0 & 1 & 1 & 1 \\
6765 & 117.819141 & 1.044834 & -0.000016 \\
10946 & 151.162392 & 0.993182 & -0.00001 \\
17711 & 193.944185 & 0.941453 & -0.000006 \\
28657 & 248.83271 & 0.889743 & -0.000004 \\
46368 & 319.249579 & 0.838128 & -0.000002 \\
75025 & 409.58232 & 0.786664 & -0.000001 \\
100000 & 475.281318 & 0.756025 & -0.000001 \\
\midrule
0 & 1 & 1 & 1 \\
6765 & 93.959956 & 2.87317 & 0.000024 \\
10946 & 117.941041 & 2.952831 & 0.000015 \\
17711 & 148.018844 & 3.033553 & 0.00001 \\
28657 & 185.734984 & 3.115385 & 0.000006 \\
46368 & 233.018627 & 3.198372 & 0.000004 \\
75025 & 292.283231 & 3.282557 & 0.000002 \\
100000 & 334.60206 & 3.333415 & 0.000002 \\
\bottomrule
\end{tabular}
\caption{Results of the physical variables shown in \cref{fig:BI_varying_gamma}, for different time instances \(t\), with \(\gamma = 0\) in the upper part, \(\gamma = -0.2\) in the central part,  and \(\gamma = 0.2\) in the lower part.}
\label{table:BI_varying_gamma}
\end{table}

Regarding the anisotropy parameter \( \beta_{nc}(t) \), its behavior follows the expected pattern: rapid growth in the initial stages, followed by stabilization toward an asymptotic value. From \cref{table:BI_varying_gamma}, we observe that the final value of \( \beta_{nc}(t) \) in \( t = 100000 \) is approximately \( 1.88 \) for \( \gamma = 0 \), \( 1.76 \) for \( \gamma = -0.2 \), and \( 3.33 \) for \( \gamma = 0.2 \). This confirms that positive values of \( \gamma \) lead to higher residual anisotropy, while negative values favor greater isotropization.

Furthermore, the time derivative \( \dot{\beta}_{nc}(t) \) decreases over time in all cases, confirming that the anisotropy tends to stabilize at late times. Again, from \cref{table:BI_varying_gamma}, we see that for \( t = 100000 \), the values obtained are \( 9.4573 \times 10^{-9} \) for \( \gamma = 0 \), \( -1 \times 10^{-6} \) for \( \gamma = -0.2 \), and \( 2 \times 10^{-6} \) for \( \gamma = 0.2 \). The lowest rate of change is observed in the commutative case, indicating that the anisotropy is nearly stabilized. In contrast, for \( \gamma = \pm 0.2 \), although the variation is small, anisotropy takes longer to reach its asymptotic state.

\subsubsection{Varying the energy density $C$ }

\begin{figure}[H]
    \centering
    \begin{subfigure}[b]{0.49\textwidth}
        \includegraphics[width=1\textwidth]{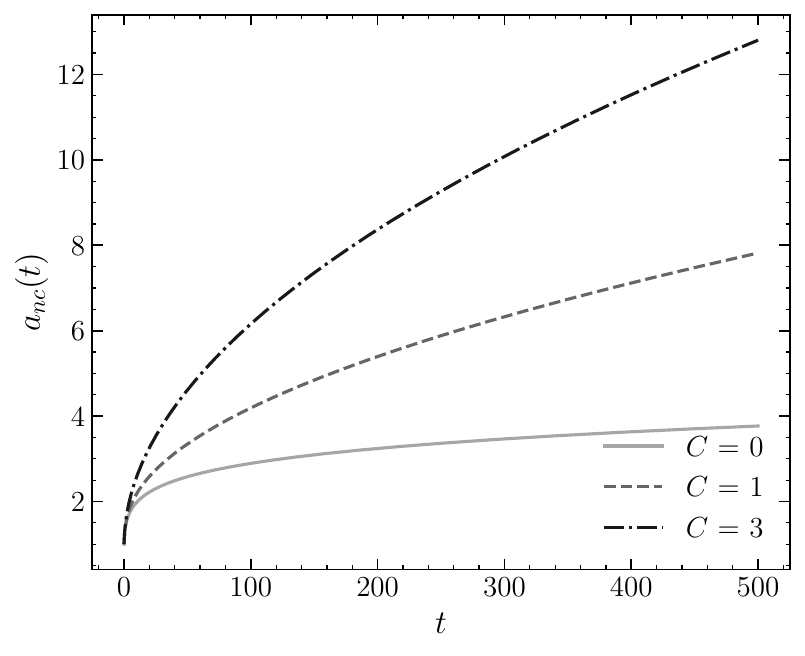} %
        \caption{ Evolution of the scale factor \(a_{nc}(t)\) as a function of time \(t\).} 
        \label{fig:BI_a_C}
    \end{subfigure}
    \hfill
    \begin{subfigure}[b]{0.49\textwidth}
        \includegraphics[width=1\textwidth]{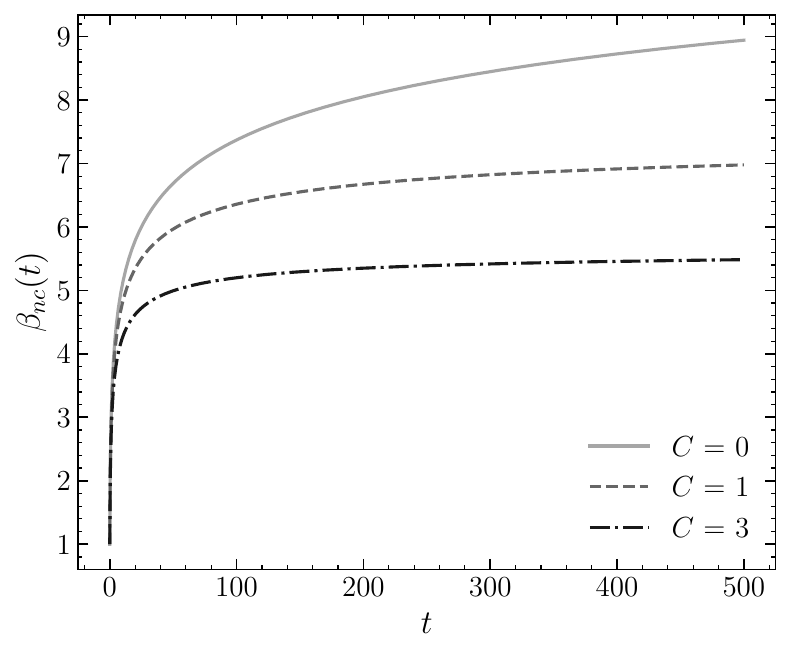} %
        \caption{Behavior of the anisotropy parameter \(\beta_{nc}(t)\) as a function of time \(t\).}
        \label{fig:BI_beta_C}
    \end{subfigure}
    \caption{\justifying The graphs (a) and (b) show the numerical solutions obtained under initial conditions: \(a_c(0) = 1\), \(\dot{a}_c(0) = 1\), \(\beta_c(0) = 1\), \(T_c(0) = 0\), and the NC parameter \(\gamma = -0.01\). For each selected value of \(C= \{0,\, 1,\, 3\}\), the corresponding value of \(\dot{\beta}_c(0)  = \{5.9746, \, 5.70785, \, 5.1066\}\) was calculated using the constraint \(H_c = 0\), while keeping the other initial conditions fixed.}
    \label{fig:BI_varying_C}
\end{figure}

\begin{table}[H]
\centering
\begin{tabular}{lS[table-format=8.0e+7] S[table-format=6.0e+6] S[table-format=6.0e+6] S[table-format=8.0e+6]}
\toprule
{\(t\)} & {\(a_{nc}(t)\)} & {\(\beta_{nc}(t)\)} & {\(\dot{\beta}_{nc}(t)\)} \\ 
\midrule
0 & 1 & 1 & 5.974579 \\
6765 & 5.825437 & 11.297205 & 0.000116 \\
10946 & 6.396436 & 11.654721 & 0.000063 \\
17711 & 7.129517 & 11.96011 & 0.000032 \\
28657 & 8.128739 & 12.198249 & 0.000015 \\
46368 & 9.52243 & 12.359923 & 0.000006 \\
75025 & 11.441132 & 12.446524 & 0.000001 \\
100000 & 12.890687 & 12.466424 & 3.213368e-07 \\
\midrule 
0 & 1 & 1 & 5.1066 \\
6765 & 45.551467 & 5.630641 & 0.000003 \\
10946 & 57.857077 & 5.63739 & 9.368731e-07 \\
17711 & 73.537506 & 5.640777 & 2.296734e-07 \\
28657 & 93.51662 & 5.641516 & -2.801714e-08 \\
46368 & 118.971422 & 5.640171 & -9.993480e-08 \\
75025 & 151.401387 & 5.637187 & -1.018942e-07 \\
100000 & 174.860954 & 5.634773 & -9.124810e-08 \\
\bottomrule
\end{tabular}
\caption{Results of the physical variables shown in \cref{fig:BI_varying_C}, for different time instances \(t\), with \(C = 0\) in the upper part and \(C = 3\) in the lower part.}
\label{table:BI_varying_C}
\end{table}

Once the effect of the parameter \( \gamma \) has been explored, the impact of the energy density parameter \( C \) on the dynamics of the system is examined. As shown in \cref{fig:BI_varying_C}, a higher value of \( C \) increases the growth of the scale factor \( a_{nc}(t) \), suggesting that the energy density of the system significantly influences the expansion rate of the universe. Regarding the evolution of the anisotropy parameter \( \beta_{nc}(t) \), it shows rapid growth in the initial stages, followed by stabilization toward an asymptotic value, indicating that the universe reaches a more stable isotropic state after some time. Furthermore, as \( C \) increases, the asymptotic value of \( \beta_{nc}(t) \) decreases.

The numerical analysis of these results, presented in \cref{table:BI_varying_C}, shows that for \( t = 100000 \), the scale factor reaches a value of \( 12.8907 \) when \( C = 0 \), and \( 174.8610 \) when \( C = 3 \), confirming that higher values of \( C \) lead to a more pronounced expansion of the scale factor \( a_{nc}(t) \). On the other hand, the anisotropy parameter \( \beta_{nc}(t) \) tends to asymptotic values of \( 12.5 \) for \( C = 0 \) and \( 5.6 \) for \( C = 3 \), highlighting the influence of the energy density on the evolution of anisotropy. This behavior indicates that anisotropy does not disappear completely, but rather converges toward finite values in the asymptotic regime. Moreover, the time derivative of the anisotropy parameter, \( \dot{\beta}_{nc}(t) \), decreases rapidly with time, reaching values on the order of \( 3.2134 \times 10^{-7} \) for \( C = 0 \) and \( -9.1248 \times 10^{-8} \) for \( C = 3 \). These results suggest that a higher energy density accelerates the isotropization process of the model, promoting a more efficient evolution toward a dynamic equilibrium state.

\subsubsection{Varying the  $a_{c}$ }

\begin{figure}[H]
    \centering
    \begin{subfigure}[b]{0.49\textwidth}
        \includegraphics[width=1\textwidth]{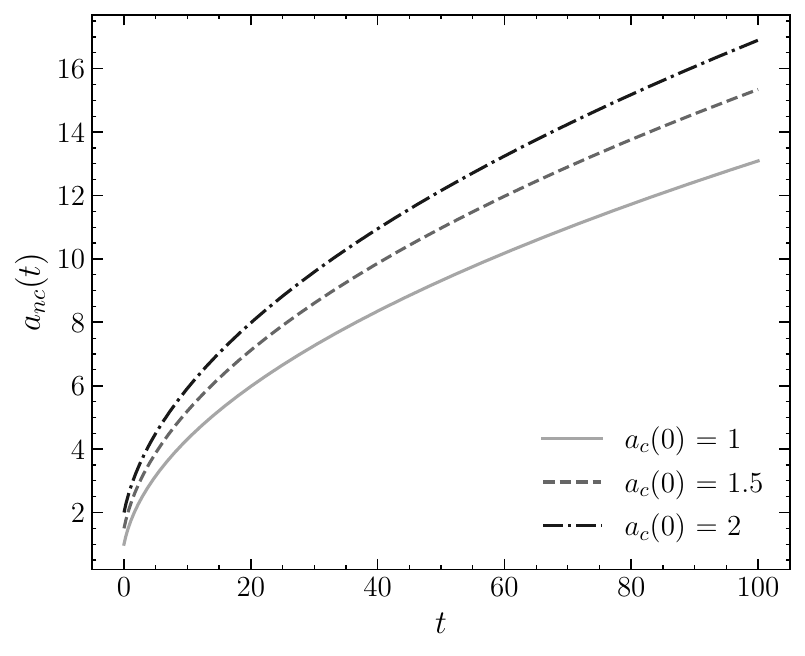} %
        \caption{ Evolution of the scale factor \(a_{nc}(t)\) as a function of time \(t\).} 
        \label{fig:BI_a_a}
    \end{subfigure}
    \hfill
    \begin{subfigure}[b]{0.49\textwidth}
        \includegraphics[width=1\textwidth]{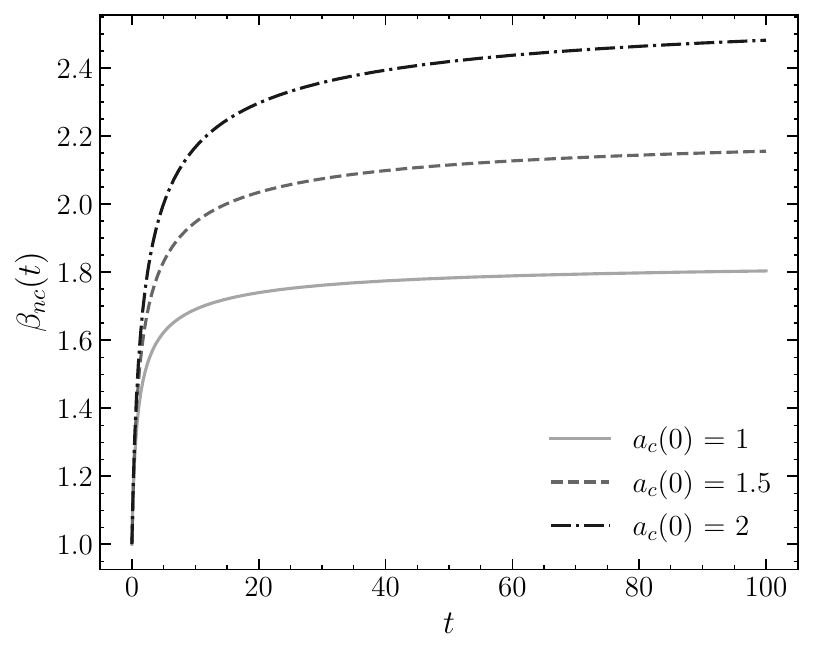} %
        \caption{Behavior of the anisotropy parameter \(\beta_{nc}(t)\) as a function of time \(t\).}
        \label{fig:BI_beta_a}
    \end{subfigure}
    \caption{\justifying The graphs (a) and (b) show the numerical solutions obtained under initial conditions: \(\dot{a}_c(0) = 1\), \(\beta_c(0) = 1\), \(\dot{\beta}_c(0) = 1\), \(T_c(0) = 0\), and the NC parameter \(\gamma = -0.01\). For each selected value of \(a_c(0) = \{1,\, 1.5,\, 2\}\), the corresponding value of \(C = \{7.2463, \, 17.163, \, 31.0318\}\) was calculated using the constraint \(H_c = 0\), while keeping the other initial conditions fixed.}
    \label{fig:BI_varying_a}
\end{figure}

\begin{table}[H]
\centering
\begin{tabular}{lS[table-format=8.0e+7] S[table-format=6.0e+6] S[table-format=6.0e+6] S[table-format=8.0e+6]} 
\toprule
{\(t\)} & {\(a_{nc}(t)\)} & {\(\beta_{nc}(t)\)} & {\(\dot{\beta}_{nc}(t)\)} \\ 
\midrule
0 & 1 & 1 & 1 \\
6765 & 106.900259 & 1.829763 & -4.087872e-07 \\
10946 & 136.011788 & 1.828233 & -3.253992e-07 \\
17711 & 173.061979 & 1.826362 & -2.364353e-07 \\
28657 & 220.215454 & 1.824224 & -1.632731e-07 \\
46368 & 280.2271 & 1.821876 & -1.092286e-07 \\
75025 & 356.603021 & 1.819363 & -7.154188e-08 \\
100000 & 411.808141 & 1.817799 & -5.516069e-08 \\
\midrule 
0 & 2 & 1 & 1 \\
6765 & 135.076439 & 2.608498 & 9.339918e-07 \\
10946 & 171.738679 & 2.611008 & 3.838477e-07 \\
17711 & 218.389065 & 2.612614 & 1.432207e-07 \\
28657 & 277.74788 & 2.613507 & 4.282327e-08 \\
46368 & 353.276429 & 2.613839 & 4.261344e-09 \\
75025 & 449.378883 & 2.613732 & -8.156041e-09 \\
100000 & 518.830117 & 2.613498 & -1.010171e-08 \\
\bottomrule
\end{tabular}
\caption{Results of the physical variables shown in \cref{fig:BI_varying_a}, for different time instances \(t\), with \(a_{c}(0) = 1\) in the upper part and \(a_{c}(0) = 2\) in the lower part.}
\label{table:BI_varying_a}
\end{table}

Continuing with the study of initial conditions, the effect of varying the initial condition \(a_c(0)\) on the evolution of the scale factor \(a_{nc}(t)\) and the anisotropy parameter \(\beta_{nc}(t)\) is now analyzed. As shown in \cref{fig:BI_varying_a}, the results indicate that increasing the initial value of \(a_c(0)\) accelerates the growth of \(a_{nc}(t)\), leading to a faster expansion of the universe. Regarding the evolution of \(\beta_{nc}(t)\), it shows rapid growth in the initial stages, followed by stabilization toward an asymptotic value. Additionally, as \(a_c(0)\) increases, this asymptotic value also increases, suggesting that the universe tends toward an isotropic stable state in later stages.

The numerical analysis of these results, presented in \cref{table:BI_varying_a}, reveals that for \( t = 100000 \), the scale factor reaches a value of \( 411.8081 \) when \( a_c(0) = 1 \), and \( 518.8301 \) when \( a_c(0) = 2 \), confirming that, by increasing the initial value of \( a_c(0) \), the expansion rate of \( a_{nc}(t) \) increases. On the other hand, the anisotropy parameter \( \beta_{nc}(t) \) tends to asymptotic values of \( 1.82 \) for \( a_c(0) = 1 \) and \( 2.61 \) for \( a_c(0) = 2 \), indicating that the anisotropy does not vanish, but instead stabilizes at finite values. Similarly, the rate of change \( \dot{\beta}_{nc}(t) \) decreases rapidly over time, reaching values on the order of \( - 5.5161 \times 10^{-8} \) for \( a_c(0) = 1 \) and \( - 1.0102 \times 10^{-8} \) for \( a_c(0) = 2 \). This suggests that, with higher values of \( a_c(0) \), the evolution toward an equilibrium state is faster, where its variation becomes practically negligible.

\subsubsection{Varying the  $\dot{a}_{c}$ }

\begin{figure}[H]
    \centering
    \begin{subfigure}[b]{0.49\textwidth}
        \includegraphics[width=1\textwidth]{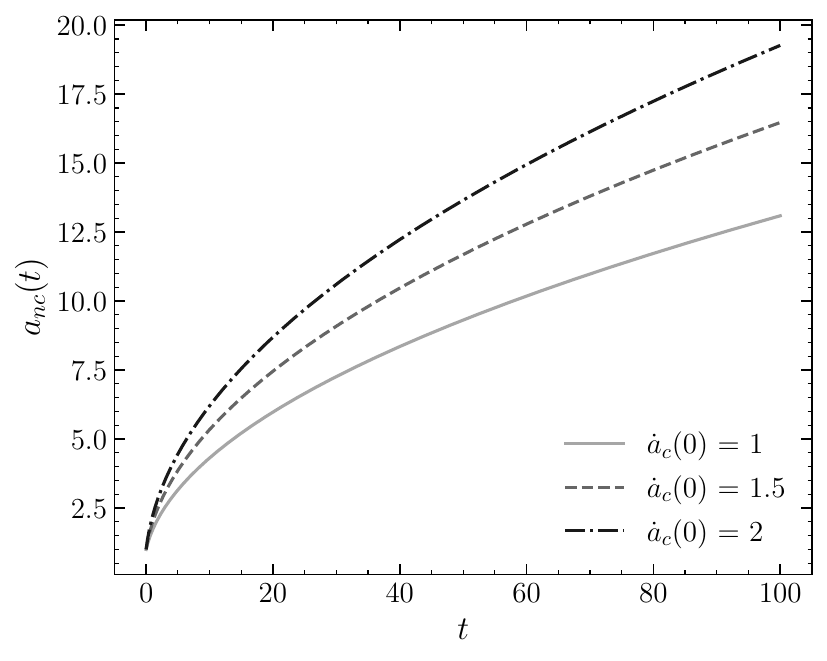} %
        \caption{ Evolution of the scale factor \(a_{nc}(t)\) as a function of time \(t\).} 
        \label{fig:BI_a_da}
    \end{subfigure}
    \hfill
    \begin{subfigure}[b]{0.49\textwidth}
        \includegraphics[width=1\textwidth]{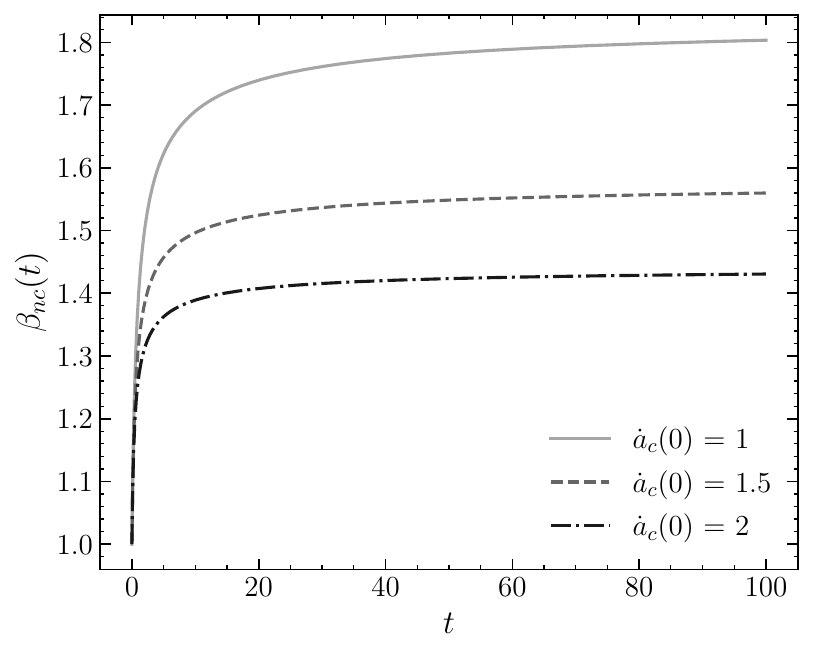} %
        \caption{Behavior of the anisotropy parameter \(\beta_{nc}(t)\) as a function of time \(t\).}
        \label{fig:BI_beta_da}
    \end{subfigure}
    \caption{\justifying The graphs (a) and (b) show the numerical solutions obtained under initial conditions: \(a_c(0) = 1\), \(\beta_c(0) = 1\), \(\dot{\beta}_c(0) = 1\), \(T_c(0) = 0\), and the NC parameter \(\gamma = -0.01\). For each selected value of \(\dot{a}_c(0) = \{1,\, 1.5,\, 2\}\), the corresponding value of \(C = \{ 7.2463, \, 15.4994, \, 26.6742\}\) was calculated using the constraint \(H_c = 0\), while keeping the other initial conditions fixed.}
    \label{fig:BI_varying_da}
\end{figure}

\begin{table}[H]
\centering
\begin{tabular}{lS[table-format=8.0e+7] S[table-format=6.0e+6] S[table-format=6.0e+6] S[table-format=8.0e+6]} 
\toprule
{\(t\)} & {\(a_{nc}(t)\)} & {\(\beta_{nc}(t)\)} & {\(\dot{\beta}_{nc}(t)\)} \\ 
\midrule
0 & 1 & 1 & 1 \\
6765 & 106.900259 & 1.829763 & -4.087872e-07 \\
10946 & 136.011788 & 1.828233 & -3.253992e-07 \\
17711 & 173.061979 & 1.826362 & -2.364353e-07 \\
28657 & 220.215454 & 1.824224 & -1.632731e-07 \\
46368 & 280.2271 & 1.821876 & -1.092286e-07 \\
75025 & 356.603021 & 1.819363 & -7.154188e-08 \\
100000 & 411.808141 & 1.817799 & -5.516069e-08 \\
\midrule 
0 & 1 & 1 & 1 \\
6765 & 158.07382 & 1.437434 & -2.638631e-07 \\
10946 & 201.102897 & 1.436501 & -1.903795e-07 \\
17711 & 255.851007 & 1.435439 & -1.309204e-07 \\
28657 & 325.509696 & 1.434277 & -8.735122e-08 \\
46368 & 414.139718 & 1.433036 & -5.711106e-08 \\
75025 & 526.907757 & 1.431733 & -3.681300e-08 \\
100000 & 608.400371 & 1.430932 & -2.817795e-08 \\
\bottomrule
\end{tabular}
\caption{Results of the physical variables shown in \cref{fig:BI_varying_da}, for different time instances \(t\), with \(\dot{a}_{c}(0) = 1\) in the upper part and \(\dot{a}_{c}(0) = 2\) in the lower part.}
\label{table:BI_varying_da}
\end{table}

In addition, the impact of varying the initial condition \( \dot{a}_c(0) \) on the evolution of \( a_{nc}(t) \) and \( \beta_{nc}(t) \) is explored. As shown in \cref{fig:BI_varying_da}, the results indicate that increasing the initial value of \( \dot{a}_c(0) \) accelerates the growth of the scale factor, promoting a faster expansion of the universe. Meanwhile, the evolution of \( \beta_{nc}(t) \) experiences rapid growth in the early stages, followed by stabilization at an asymptotic value. Additionally, as \( \dot{a}_c(0) \) increases, this asymptotic value decreases, and the universe tends toward an isotropic state in the long term.

The numerical analysis of these results, presented in \cref{table:BI_varying_da}, indicates that for \( t = 100000 \), the scale factor reaches a value of \( 411.8081 \) when \( \dot{a}_c(0) = 1 \) and \( 608.4004 \) when \( \dot{a}_c(0) = 2 \), confirming that a higher initial value of \( \dot{a}_c(0) \) increases the expansion rate of the scale factor. On the other hand, the anisotropy parameter \( \beta_{nc}(t) \) tends to asymptotic values of \( 1.82 \) for \( \dot{a}_c(0) = 1 \) and \( 1.43 \) for \( \dot{a}_c(0) = 2 \), which means that increasing the value of \( \dot{a}_c(0) \) decreases its asymptotic value. Furthermore, the rate of change \( \dot{\beta}_{nc}(t) \) decreases rapidly over time, reaching values on the order of \( - 5.5161 \times 10^{-8} \) for \( \dot{a}_c(0) = 1 \) and \( - 2.8178 \times 10^{-8} \) for \( \dot{a}_c(0) = 2 \). This indicates that as \( \dot{a}_c(0) \) increases, the anisotropy evolves more rapidly toward an equilibrium state.

\subsubsection{Varying the  $\beta_{c}$ }

\begin{figure}[H]
    \centering
    \begin{subfigure}[b]{0.49\textwidth}
        \includegraphics[width=1\textwidth]{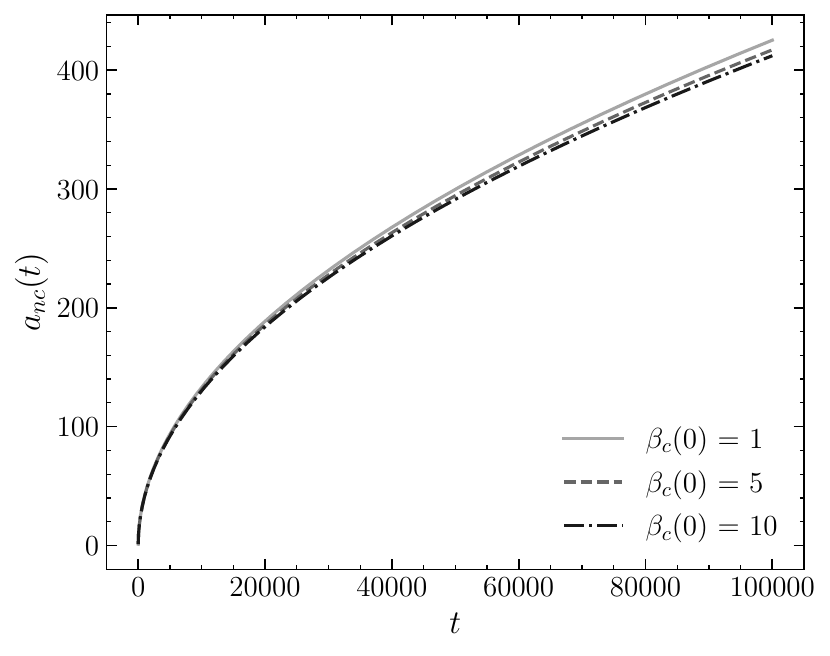} %
        \caption{ Evolution of the scale factor \(a_{nc}(t)\) as a function of time \(t\).} 
        \label{fig:BI_a_b}
    \end{subfigure}
    \hfill
    \begin{subfigure}[b]{0.49\textwidth}
        \includegraphics[width=1\textwidth]{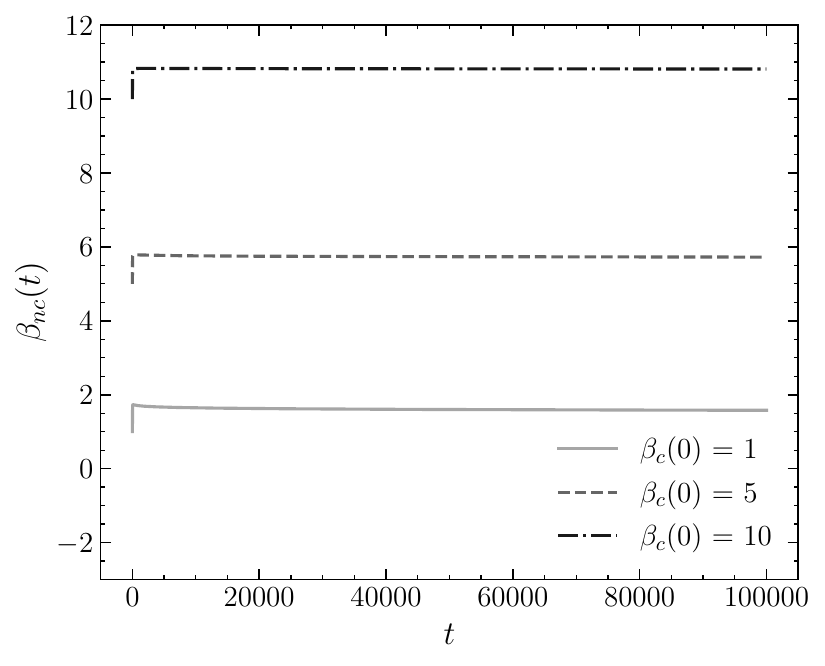} %
        \caption{Behavior of the anisotropy parameter \(\beta_{nc}(t)\) as a function of time \(t\).}
        \label{fig:BI_beta_b}
    \end{subfigure}
    \caption{\justifying The graphs (a) and (b) show the numerical solutions obtained under initial conditions: \(a_c(0) = 1\), \(\dot{a}_c(0) = 1\), \(\dot{\beta}_c(0) = 1\), \(T_c(0) = 0\), and the NC parameter \(\gamma = -0.05\). For each selected value of \(\beta_c(0) = \{1,\, 5,\, 10\}\), the corresponding value of \(C = \{7.0155, \, 102.9870, \, 2920.6700\}\) was calculated using the constraint \(H_c = 0\), while keeping the other initial conditions fixed.}
    \label{fig:BI_varying_b}
\end{figure}

Subsequently, the effect of varying the initial condition \( \beta_c(0) \) on the evolution of the system is examined. As shown in \cref{fig:BI_varying_b}, an increase in the initial value of \( \beta_c(0) \) does not favor the growth of the scale factor \( a_{nc}(t) \), it slightly reduces its expansive character. Regarding the dynamics of \( \beta_{nc}(t) \), rapid growth is observed in the early stages, followed by stabilization toward an asymptotic value. Additionally, as \( \beta_c(0) \) increases, the value of this asymptote also increases, suggesting that although anisotropy does not vanish, the universe evolves toward an isotropic regime at later times.

\begin{table}[H]
\centering
\begin{tabular}{lS[table-format=8.0e+7] S[table-format=6.0e+6] S[table-format=6.0e+6] S[table-format=8.0e+6]}
\toprule
{\(t\)} & {\(a_{nc}(t)\)} & {\(\beta_{nc}(t)\)} & {\(\dot{\beta}_{nc}(t)\)} \\ 
\midrule
0 & 1 & 1 & 1 \\
6765 & 109.209163 & 1.655316 & -0.000004 \\
10946 & 139.22317 & 1.642023 & -0.000003 \\
17711 & 177.496848 & 1.628388 & -0.000002 \\
28657 & 226.302709 & 1.614493 & -0.000001 \\
46368 & 288.538477 & 1.600401 & -6.352428e-07 \\
75025 & 367.899073 & 1.58616 & -3.961225e-07 \\
100000 & 425.348972 & 1.577602 & -2.984159e-07 \\
\midrule 
0 & 1 & 10 & 1 \\
6765 & 106.986231 & 10.825374 & -5.154552e-07 \\
10946 & 136.129134 & 10.823495 & -3.919271e-07 \\
17711 & 173.221495 & 10.821273 & -2.778525e-07 \\
28657 & 220.431491 & 10.818781 & -1.890187e-07 \\
46368 & 280.5187 & 10.816077 & -1.252127e-07 \\
75025 & 356.995411 & 10.813207 & -8.145552e-08 \\
100000 & 412.276005 & 10.811429 & -6.261101e-08 \\
\bottomrule
\end{tabular}
\caption{Results of the physical variables shown in \cref{fig:BI_varying_b}, for different time instances \(t\), with \(\beta_{c}(0) = 1\) in the upper part and \(\beta_{c}(0) = 10\) in the lower part.}
\label{table:BI_varying_b}
\end{table}

The corresponding numerical analysis, presented in \cref{table:BI_varying_b}, indicates that for \( t = 100000 \), the scale factor reaches a value of \( 425.3490 \) when \( \beta_c(0) = 1 \) and \( 412.2760 \) when \( \beta_c(0) = 10 \), which confirms that a higher initial anisotropy does not increase the expansion of \( a_{nc}(t) \). On the other hand, the anisotropy parameter \( \beta_{nc}(t) \) tends to asymptotic values of \( 1.6 \) for \( \beta_c(0) = 1 \) and \( 10.8 \) for \( \beta_c(0) = 10 \), indicating that the anisotropy does not disappear but instead stabilizes at finite values. Finally, the time derivative \( \dot{\beta}_{nc}(t) \) decreases rapidly, reaching values of the order of $ - 2.9842 \times 10^{-7}$  for $ \beta_c(0) = 1 $ and $ - 6.2611 \times 10^{-8} $ for $\beta_c(0) = 10$, which means that as the value of $\beta_c(0)$ increases, the isotropization process becomes faster.

\subsubsection{Varying the $\dot{\beta}_{c}$ }

\begin{figure}[H]
    \centering
    \begin{subfigure}[b]{0.49\textwidth}
        \includegraphics[width=1\textwidth]{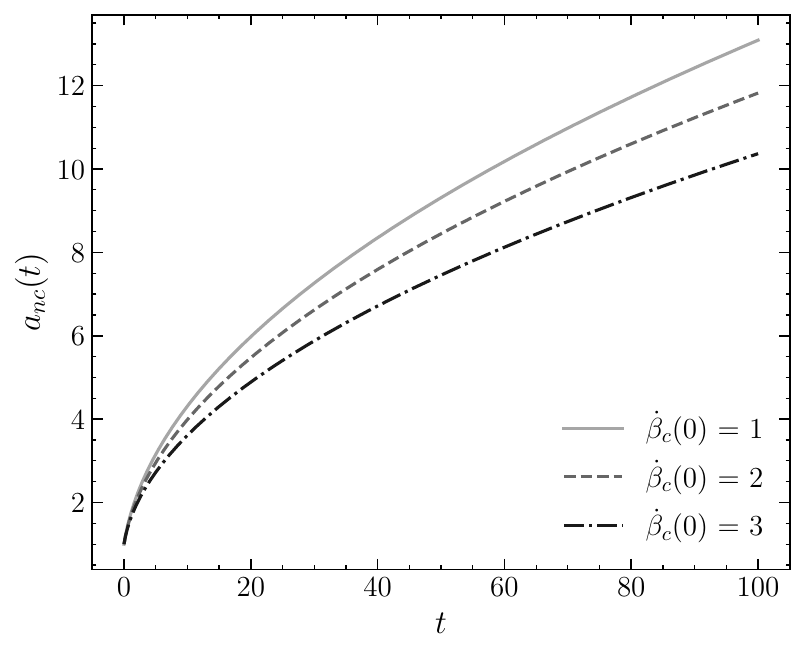} %
        \caption{ Evolution of the scale factor \(a_{nc}(t)\) as a function of time \(t\).} 
        \label{fig:BI_a_db}
    \end{subfigure}
    \hfill
    \begin{subfigure}[b]{0.49\textwidth}
        \includegraphics[width=1\textwidth]{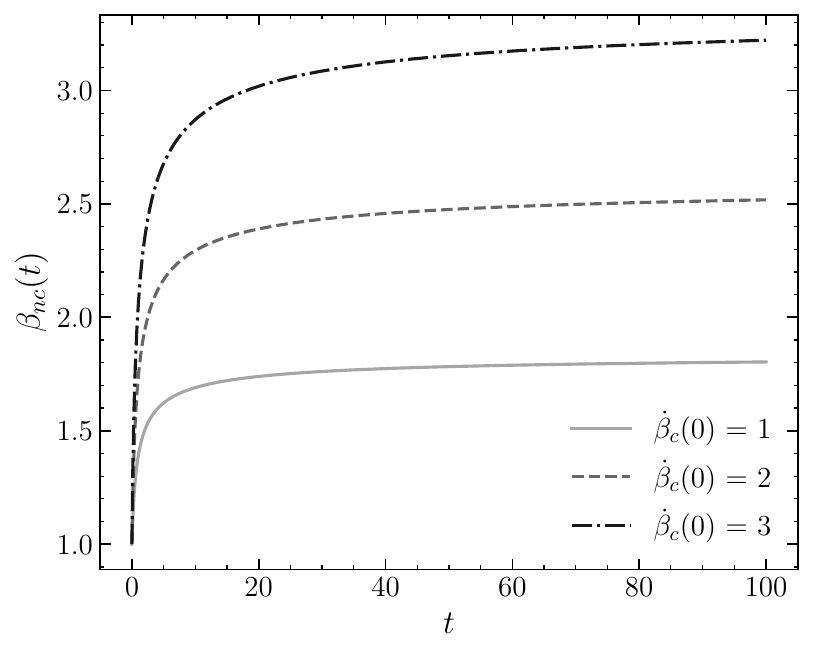} %
        \caption{Behavior of the anisotropy parameter \(\beta_{nc}(t)\) as a function of time \(t\).}
        \label{fig:BI_beta_db}
    \end{subfigure}
    \caption{\justifying The graphs (a) and (b) show the numerical solutions obtained under initial conditions: \(a_c(0) = 1\), \(\dot{a}_c(0) = 1\), \(\beta_c(0) = 1\), \(T_c(0) = 0\), and the NC parameter \(\gamma = -0.01\). For each selected value of \(\dot{\beta}_c(0) = \{1,\, 2,\, 3\}\), the corresponding value of \(C = \{7.2463, \, 7.7250, \, 7.2298\}\) was calculated using the constraint \(H_c = 0\), while keeping the other initial conditions fixed.}
    \label{fig:BI_varying_db}
\end{figure}

Finally, the effect of varying the initial condition \( \dot{\beta}_c(0) \) on the evolution of the system is studied. As shown in \cref{fig:BI_varying_db}, the results indicate that a higher value of \( \dot{\beta}_c(0) \) slows the growth of the scale factor \( a_{nc}(t) \), although without modifying its expansive character. Regarding the evolution of \( \beta_{nc}(t) \), it shows a rapid increase in the early stages, followed by stabilization toward an asymptotic value. Furthermore, it is observed that as \( \dot{\beta}_c(0) \) increases, the value of this asymptote also increases.

\begin{table}[H]
\centering
\begin{tabular}{lS[table-format=8.0e+7] S[table-format=6.0e+6] S[table-format=6.0e+6] S[table-format=8.0e+6]}
\toprule
{\(t\)} & {\(a_{nc}(t)\)} & {\(\beta_{nc}(t)\)} & {\(\dot{\beta}_{nc}(t)\)} \\ 
\midrule
0 & 1 & 1 & 1 \\
6765 & 106.900259 & 1.829763 & -4.087872e-07 \\
10946 & 136.011788 & 1.828233 & -3.253992e-07 \\
17711 & 173.061979 & 1.826362 & -2.364353e-07 \\
28657 & 220.215454 & 1.824224 & -1.632731e-07 \\
46368 & 280.2271 & 1.821876 & -1.092286e-07 \\
75025 & 356.603021 & 1.819363 & -7.154188e-08 \\
100000 & 411.808141 & 1.817799 & -5.516069e-08 \\
\midrule 
0 & 1 & 1 & 3 \\
6765 & 82.988003 & 3.346053 & 4.132038e-07 \\
10946 & 105.54412 & 3.346802 & 3.778496e-08 \\
17711 & 134.256229 & 3.346532 & -8.239524e-08 \\
28657 & 170.803925 & 3.345462 & -1.023013e-07 \\
46368 & 217.325147 & 3.343761 & -8.818094e-08 \\
75025 & 276.541235 & 3.341566 & -6.661610e-08 \\
100000 & 319.348258 & 3.340064 & -5.444791e-08 \\
\bottomrule
\end{tabular}
\caption{Results of the physical variables shown in \cref{fig:BI_varying_db}, for different time instances \(t\), with \(\dot{\beta}(0) = 1\) in the upper part and \(\dot{\beta}(0) = 3\) in the lower part.}
\label{table:BI_varying_db}
\end{table}

The corresponding numerical analysis, presented in \cref{table:BI_varying_db}, shows that for \( t = 100000 \), the scale factor reaches a value of \( 411.8081 \) when \( \dot{\beta}_c(0) = 1 \), and \( 319.3483 \) when \( \dot{\beta}_c(0) = 3 \), confirming that a higher initial value of \(\dot{\beta}_c(0) \) reduces the expansion rate of the scale factor \( a_{nc}(t) \). On the other hand, the anisotropy parameter \( \beta_{nc}(t) \) tends to asymptotic values of \( 1.8 \) for \( \dot{\beta}_c(0) = 1 \) and \( 3.3 \) for \( \dot{\beta}_c(0) = 3 \), which indicates that the anisotropy does not disappear but instead stabilizes at finite values, evolving toward an equilibrium state in later times. Furthermore, the time derivative \( \dot{\beta}_{nc}(t) \) decreases rapidly over time, reaching values on the order of \( - 5.5161 \times 10^{-8} \) for \( \dot{\beta}_c(0) = 1 \) and \( - 5.4448\times 10^{-8}\) for \(\dot{\beta}_c(0) = 3\). 

\subsection{Influence of spatial curvature on NC dynamics}
Here, we compare how different spatial curvatures, specifically flat (\( \kappa = 0 \)) and open (\( \kappa = -1 \)) geometries, affect the evolution of the universe under the influence of a NC parameter \( \gamma \). Both negative and positive values of \( \gamma \) are considered for the scale factor and the anisotropy parameter.

\subsubsection{Case of negative NC parameter ($\gamma < 0$) }
Following the detailed analysis of the cases with negative $\kappa = -1$ and zero $\kappa = 0$ spatial curvature, presented in subsections \ref{sec:level 4.1} and \ref{sec:level 4.2}, respectively, we now proceed to a direct comparison between both scenarios. This comparison focuses on the regime $\gamma < 0$, with the aim of clearly visualizing the dynamical differences induced by spatial curvature. In particular, the evolution of both the scale factor and the anisotropy parameter is evaluated together.

\begin{figure}[H]
	\centering
	\begin{subfigure}[b]{0.49\textwidth}
		\includegraphics[width=1\textwidth]{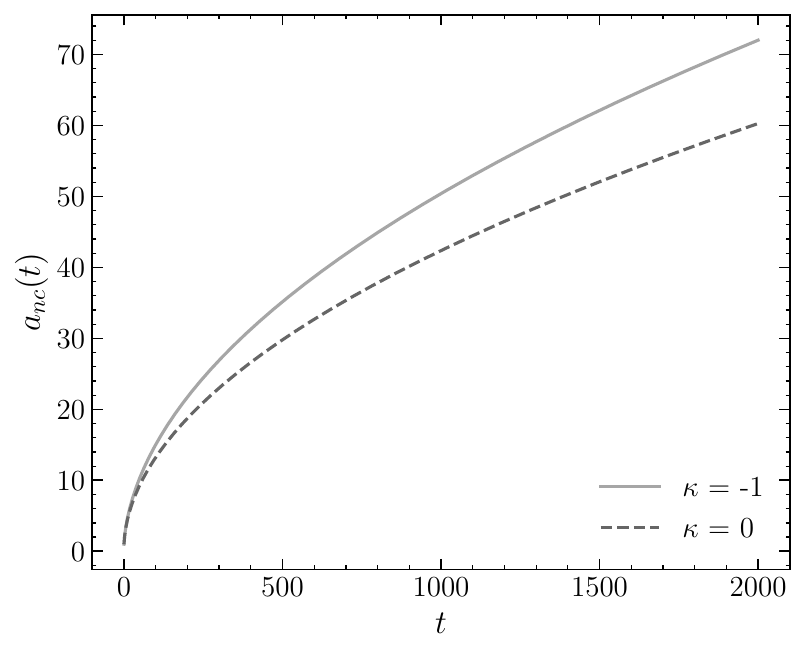} %
		\caption{ Evolution of the scale factor \(a_{nc}(t)\) as a function of time \(t\).} 
		\label{fig:CompN_a_db}
	\end{subfigure}
	\hfill
	\begin{subfigure}[b]{0.49\textwidth}
		\includegraphics[width=1\textwidth]{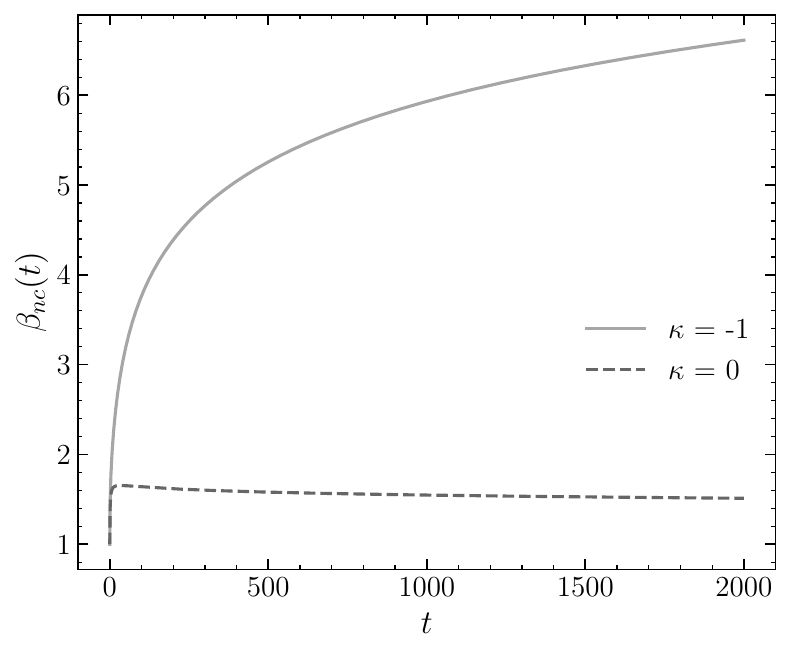} %
		\caption{Behavior of the anisotropy parameter \(\beta_{nc}(t)\) as a function of time \(t\).}
		\label{fig:CompN_beta_db}
	\end{subfigure}
	\caption{\justifying The graphs (a) and (b) show the numerical solutions obtained under initial conditions: \(a_c(0) = 1\), \(\dot{a}_c(0) = 1\), \(\beta_c(0) = 1\), \(T_c(0) = 0\), \(\dot{\beta}_c(0) = 1\) and the NC parameter \(\gamma = -0.1\). For negative ($\kappa = -1$) and flat ($\kappa = 0$) geometry, the corresponding value of \(C = \{6.0104, \, 6.7269\}\) was calculated using the constraint \(H_c = 0\), while keeping the initial conditions fixed.}
	\label{fig:CompN}
\end{figure}

The results in \cref{fig:CompN} indicate that the inclusion of negative curvature improves the universe's expansion and substantially modifies the anisotropic behavior. The scale factor $a_{nc}(t)$ exhibits accelerated growth in both cases, though with a significantly higher rate when $\kappa = -1$, suggesting that open curvature acts as an amplifier of the expansive effect generated by noncommutativity. Meanwhile, the anisotropy parameter $\beta_{nc}(t)$, which tends to stabilize in the flat case, continues to grow under negative curvature, even at a decreasing rate. This behavior suggests that an open geometry not only supports expansion but also delays the isotropization of the universe, extending the presence of anisotropies in the $\gamma < 0$ regime.

\subsubsection{Case of  positive NC parameter ($\gamma > 0$) }
\begin{figure}[H]
	\centering
	\begin{subfigure}[b]{0.49\textwidth}
		\includegraphics[width=1\textwidth]{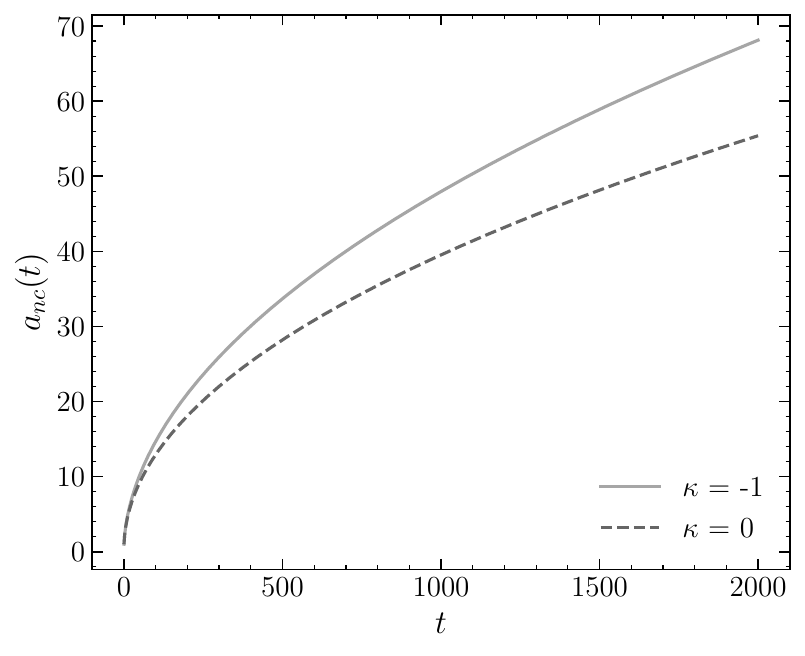} %
		\caption{ Evolution of the scale factor \(a_{nc}(t)\) as a function of time \(t\).} 
		\label{fig:CompP_a_db}
	\end{subfigure}
	\hfill
	\begin{subfigure}[b]{0.49\textwidth}
		\includegraphics[width=1\textwidth]{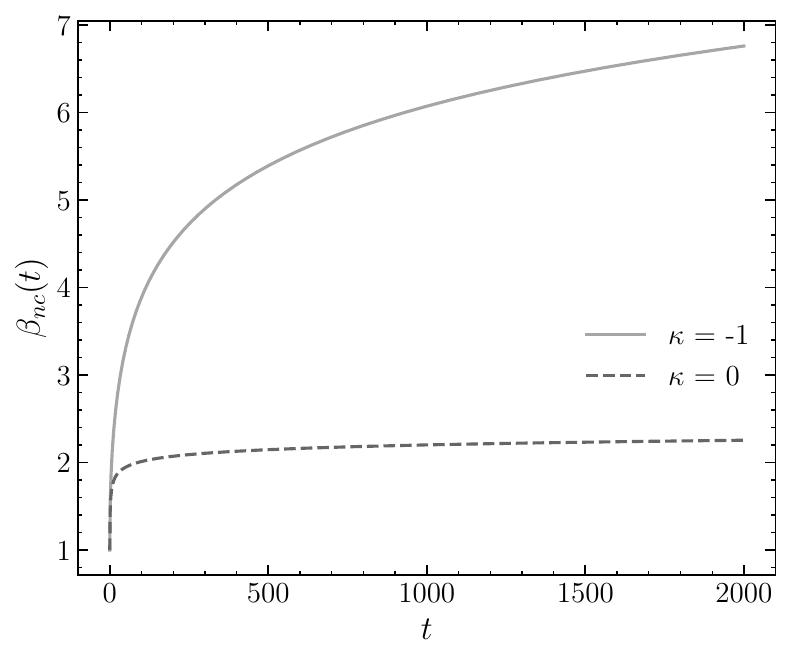} %
		\caption{Behavior of the anisotropy parameter \(\beta_{nc}(t)\) as a function of time \(t\).}
		\label{fig:CompP_beta_db}
	\end{subfigure}
	\caption{\justifying The graphs (a) and (b) show the numerical solutions obtained under initial conditions: \(a_c(0) = 1\), \(\dot{a}_c(0) = 1\), \(\beta_c(0) = 1\), \(T_c(0) = 0\), \(\dot{\beta}_c(0) = 1\) and the NC parameter \(\gamma = 0.1\). For negative ($\kappa = -1$) and flat ($\kappa = 0$) geometry, the corresponding value of \(C = \{7.1645, \, 7.8811\}\) was calculated using the constraint \(H_c = 0\), while keeping the initial conditions fixed.}
	\label{fig:CompP}
\end{figure}

After analyzing the dynamical behavior of the model for $\gamma > 0$ in \cref{fig:CompP}, it is observed that the inclusion of negative spatial curvature ($\kappa = -1$) continues to exert an amplifying effect on the expansion of the universe, although to a smaller extent than in the $\gamma < 0$ regime. The scale factor $a_{nc}(t)$ grows in both scenarios, but reaches significantly higher values in the presence of negative curvature, suggesting that this type of geometry improves the expansion rate even when the NC parameter tends to reduce it. On the other hand, the anisotropy parameter $\beta_{nc}(t)$ exhibits a notably different behavior in each case: while in the flat model ($\kappa = 0$) anisotropy stabilizes at late times, in the $\kappa = -1$ case it continues to grow, reaching much higher values. This result indicates that the combination of open geometry and noncommutativity with $\gamma > 0$ obstructs the isotropization process, prolonging the presence of anisotropies in the universe over time.

\subsection{For the case of $\kappa= +1$}

In contrast to the solutions of the system of equations \eqref{approx_dda}-\eqref{approx_Hc} for \(\kappa = 0\) and \(\kappa = -1\), corresponding to the BI and BIII cosmological models, respectively, the solution for \(\kappa = 1\), associated with the Kantowski-Sachs cosmological model, exhibits dynamic features of gravitational collapse. In particular, this model does not show sustained expansive behavior in its scale factor, and its anisotropy parameter does not tend toward isotropy in the asymptotic regime during the radiation era. For these reasons, the numerical analysis of the solution corresponding to \(\kappa = 1\) falls outside the scope and interest of the present work, which focuses on models that tend toward isotropic and expansive evolution.

\section{Estimations of the NC Parameter $\gamma$ value with observational data}

In this section, we estimate the value of the parameter \(\gamma\) using current observational data. To this end, we employ the NC Hamiltonian given in equation \eqref{NCSuperHamiltonianMatter} of the model corresponding to \(\kappa = -1\), which describes an open universe in the context of Bianchi III. We assume the conditions of isotropy and homogeneity for the current Universe, implying that \(\beta_{nc} \to \text{constant}\) and its associated canonical momentum \(P_{\beta_{nc}} \to 0\). Within this framework, we consider that the observed accelerated expansion is entirely attributed to the presence of \(\gamma\), which plays a crucial role in the dynamics of the Universe. Under these conditions, the NC Hamiltonian of the model reduces to
\begin{equation}\label{Hamiltonian_NC_actual}
     H_{nc} =   -\kappa a_{nc} -\dfrac{P_{a_{nc}}^{2}}{16a_{nc}} +  a_{nc}^{-3\alpha} P_{T_{nc}}.
\end{equation}
\noindent Unlike the previous sections, in this section we work directly with the physical variables, that is, the NC variables. This will allow us to determine the isotropic scale factor \(a_{nc}\) as a function of time.

Starting from the NC Hamiltonian in equation \eqref{Hamiltonian_NC_actual}, we calculate the dynamic equations using the deformed Poisson brackets described in equations \eqref{PoissonBracket a}-\eqref{PoissonBracket c}. These modified brackets allow us to obtain the time evolution of the system's variables under the influence of non-commutativity. The resulting dynamic equations are:
\begin{equation}\label{MovmentEq_actual a}
     \dot{a}_{nc} = \{a_{nc},\, H_{nc}\} = -\frac{P_{a_{nc}}}{8 a_{nc}} + \gamma a_{nc}^{-3 \alpha},
\end{equation}
\begin{equation}\label{MovmentEq_actual b}
    \dot{P}_{a_{nc}} = \{P_{a_{nc}},\, H_{nc}\} = 3 \alpha P_{T_{nc}}a_{nc}^{-3 \alpha -1}-\frac{P_{a_{nc}}^2}{16 a_{nc}^2}+\kappa ,
\end{equation}
\begin{equation}\label{MovmentEq_actual c}
     \dot{T}_{nc} = \{T_{nc},\, H_{nc}\} =  a_{nc}^{-3 \alpha }-\frac{\gamma  P_{a_{nc}}}{8 a_{nc}} ,
\end{equation}
\begin{equation}\label{MovmentEq_actual d}
    \dot{P}_{T_{nc}} = \{P_{T_{nc}},\, H_{nc}\} = 3 \alpha  \gamma P_{T_{nc}} a_{nc}^{-3 \alpha -1}-\frac{\gamma P_{a_{nc}}^2}{16 a_{nc}^2}+\kappa\gamma .
\end{equation}

We can observe that equations \eqref{MovmentEq_actual b} and \eqref{MovmentEq_actual d} are proportionally connected through the parameter \(\gamma\). This connection is
\begin{equation}\label{Pa_conction_PT}
    \dot{P}_{T_{nc}} = \gamma \dot{P}_{a_{nc}}.
\end{equation}

\noindent This expression can be easily integrated, resulting in
\begin{equation}\label{Pa_integrated_PT}
    P_{T_{nc}} = \gamma P_{a_{nc}} + C_{f},
\end{equation}
\noindent where \( C_f \) is a constant related to the energy density of the fluid.

Since we are interested in deriving a dynamic equation for the scale factor \(a_{nc}\), we can use equation \eqref{MovmentEq_actual a} to solve for \(P_{a_{nc}}\), which leads us to
\begin{equation}\label{Pa_despejado}
    P_{a_{nc}} = -8 a_{nc} \left(\dot{a}_{nc} - \gamma  a_{nc}^{-3 \alpha } \right).
\end{equation}

It is assumed that the total energy of the system is constant and does not change over time, which leads to equation \eqref{Hamiltonian_NC_actual} being expressed as
\begin{equation}\label{Hamiltonian_NC_Igual_Cero}
     -\kappa a_{nc} -\dfrac{P_{a_{nc}}^{2}}{16a_{nc}} +  a_{nc}^{-3\alpha} P_{T_{nc}}=0.
\end{equation}
\noindent Then, substituting the values of \(P_{T_{nc}}\) and \(P_{a_{nc}}\) given in equations \eqref{Pa_integrated_PT} and \eqref{Pa_despejado}, we obtain
\begin{equation}\label{Hnc_Igual_Cero}
     4 \gamma ^2 a_{nc}^{1-6 \alpha } + C_{f} a_{nc}^{-3 \alpha }-4 a_{nc} \dot{a}_{nc}^2-\kappa a_{nc} = 0.
\end{equation}
\noindent Furthermore, in the matter-dominated phase, the constant \(\alpha = 0\), and the accelerated expansion is entirely due to the presence of the NC parameter \(\gamma\). Thus, solving for \(\dot{a}_{nc}\) from the previous equation with \(\alpha = 0\), we obtain
\begin{equation}\label{a_nc despejado de H_nc}
     \dot{a}_{nc} = \left(  \frac{C_{f}}{4 a_{nc}} - \frac{\kappa}{4} +  \gamma^{2} \right)^{1/2}.
\end{equation}

From equation \eqref{a_nc despejado de H_nc}, it is possible to determine a relationship between the NC scale factor \(a_{nc}\) and the age of the universe $t$. This is done by integrating the above equation from \( t^{\text{eq}} \) to \( t \) and 
\( a_{nc}^{\text{eq}} \) to \( a_{nc} \), where the initial conditions \( t^{\text{eq}} \) and \( a_{nc}^{\text{eq}} \) represent, respectively, the age and the scale factor of the universe when the densities of matter and radiation were equal. The resulting relationship is
\begin{equation}\label{tiempo}
    	\begin{aligned}
    	t = g^{-3/2} F(a_{nc}, C_{f}, \gamma) - g^{-3/2} F(a_{nc}^{eq}, C_{f}, \gamma) + t^{eq},
    	\end{aligned}
\end{equation}
\noindent with
\begin{equation}\label{F_funcion}
    	\begin{aligned}
    	F(a_{nc}, C_{f}, \gamma) = \sqrt{A+g\,a_{nc}}\sqrt{g\,a_{nc}}-A\, \ln\left(\frac{\sqrt{g\,a_{nc}}+\sqrt{A+g\,a_{nc}}}{\sqrt{A}}\right) ,
    	\end{aligned}
\end{equation}
\noindent with \(A = 4C_{f}\) and \(g = -\kappa/4 + \gamma^{2}\). 

By comparing the Hubble function in the standard Friedmann–Robertson–Walker (FRW) cosmological model, given by \( \dot{a} = H_0 (\Omega_m/a -\kappa c^{2}/H_0^{2} )^{1/2} \), with equation \eqref{a_nc despejado de H_nc}, we identify \( C_f = 4 \Omega_m H_0^2 \) and \( g = \Omega_k H_0^2/4 + \gamma^2 \), \(\Omega_{k}\equiv -\kappa c^{2}/H_{0}^{2}\) ($c$ is the speed of light). Here, \( \Omega_m = 0.315 \) is the density parameter of matter, \( \Omega_k = 0.001 \) is the curvature parameter, and \( H_0 =(100 \,h)\, \text{km/s/Mpc} = 67.4 \, \text{km/s/Mpc} \) is the Hubble constant. These values of the cosmological parameters were extracted from the 2018 Planck results \cite{aghanim2020planck}.

To estimate the value of the NC parameter \(\gamma\), it is necessary to provide the value of the scale factor \(a_{nc}^{eq} = 4.15 \times 10^{-5}/(\Omega_{m}h^{2})\) and the age of the universe \(t^{eq}\) corresponding to the epoch when the densities of matter and radiation were equal \cite{dodelson2020modern}. Additionally, the values of the scale factor at another moment in the universe's history, \(a_{nc}\), and the age of the universe \(t(a_{nc})\) associated with that scale factor must also be provided. These will be determined using a ``cosmological calculator" for an open-curvature universe (\(\kappa = -1\)) \cite{wright2006cosmology}. Finally, substituting these values into the equation \eqref{tiempo}, the estimated value of the NC parameter can be obtained for the model. In \cref{table:gamma estimate}, we present the estimated values of \(\gamma\).

\begin{table}[H]
	\begin{center}
		\sisetup{print-zero-exponent=false}
		\begin{tabular}{|S[table-format=2.3e+5]|*{6}{S[table-format =2.3e+5]|}}\toprule 
		     { Scale factor ($a_{nc}$)}    & {Age of the Universe in years (\(t\))} & { NC parameter (\(\gamma\))}\\ \midrule
			1.00 & 11.65900e9 & 5.710e-11 \\
            0.90 & 10.22000e9 & 5.712e-11\\
            0.80 & 8.80800e9 &  5.713e-11 \\
            0.70 & 7.42800e9 &  5.715e-11\\
            0.60 & 6.08700e9 &  5.718e-11 \\
            0.50 & 4.79400e9 &  5.722e-11 \\
            0.40 & 3.56300e9 &  5.726e-11 \\
            0.30 & 2.41200e9 &  5.739e-11 \\
            0.20 & 1.37400e9 &  5.779e-11 \\
            0.10 & 5.11000e8 &  5.926e-11 \\
            0.05 & 1.85000e8 &  6.562e-11 \\
            0.01 & 1.65000e6 &  6.028e-9 \\
			\bottomrule
		\end{tabular}
		\caption{The scale factor (\(a_{nc}\)), the age of the Universe (\(t\)) in years, and the NC parameter (\(\gamma\)) for a cosmological model. The data illustrates the relationship between cosmic expansion and the corresponding parameters at different epochs.}
		\label{table:gamma estimate}
	\end{center}
\end{table}
In \cref{table:gamma estimate}, it is observed that as the scale factor decreases to smaller values, the age of the Universe also decreases, reaching only \(1.65 \times 10^6\) years for \(a_{nc} = 0.01\). Simultaneously, the parameter \(\gamma\), which has remained nearly constant in recent epochs (\(a_{nc} \geq 0.10\)), experiences a significant increase in the early Universe, reaching a value of \(6.028 \times 10^{-9}\). This behavior suggests that the NC effects represented by \(\gamma\) are more significant during the early stages of the evolution of the Universe and decrease as the Universe expands.

\newpage
\section{Discussion and Conclusions}

The results obtained in this study provide a new perspective on the evolution of the universe in Bianchi I and Bianchi III cosmological models by incorporating a NC parameter \( \gamma \). This parameter, which extends the classical geometry of phase space into a NC context, has a significant impact on the dynamics of the universe’s expansion. It is observed that, for \( \gamma < 0 \), the expansion of the universe is increased compared to the commutative case (\( \gamma = 0 \)), while for \( \gamma > 0 \), the expansion is slightly reduced. This behavior can be attributed to the correction introduced by the term \( \left(\frac{\gamma}{2}\right) T_c \) to the scale factor \( a_{nc} \), which effectively modifies the value of the NC scale factor. Although this term does not represent a force in the physical sense, it reflects how the structure of the NC phase space alters the geometry and, therefore, the dynamics of the expansion. Thus, in the context of the model, the negative values of \( \gamma \) tend to amplify the expansion, while positive values attenuate it, suggesting that the effects of non-commutativity can significantly influence cosmic evolution without resorting to additional exotic energy terms.

This finding suggests that the NC model could be a viable alternative to the cosmological constant \( \Lambda \), as both parameters have similar effects on the expansion of the universe, although with important differences in their dynamics. In particular, the presence of \( \gamma \) significantly affects the temporal evolution of the anisotropy \( \beta_{nc}(t) \). In the flat curvature model (Bianchi I), \( \gamma \) favors the isotropization of the universe, which is comparable to the behavior observed with a positive cosmological constant (\( \Lambda > 0 \)). However, in the negative curvature model (Bianchi III), the anisotropy persists for a longer time, which could be explained by the influence of negative curvature, which slows down the isotropization of the universe.

The increase in energy density \( C \) also shows a more accelerated expansion of the universe, which is consistent with standard cosmological theory. Moreover, the estimated values of \( \gamma \) for a homogeneous and isotropic universe indicate that this parameter was more relevant in the early stages of the universe, matching the values predicted by previous studies. This behavior underscores the dynamic nature of non-commutativity in the cosmological context, especially in the early phases of the universe.

Regarding the estimation of \( \gamma \), it was found that the current values of this parameter are small, with \( a_{nc} = 1 \), indicating that the influence of non-commutativity has diminished over time in a homogeneous and isotropic universe. However, the values of \( \gamma \) were considerably higher during the early stages of the universe, as shown in the simulated results. This behavior is expected, as non-commutativity should have had greater relevance in the early moments when the universe was denser and spatial scales were smaller. This `evolution' of $\gamma$ agrees with the values estimated by other studies, reinforcing the validity of this model in describing the evolution of the early universe \cite{gil4,gil7,gil8}.

In conclusion, non-commutativity provides a powerful tool for explaining cosmic acceleration and the early evolution of the universe, opening new avenues for understanding phenomena such as accelerated expansion and isotropization in different cosmological models. This study suggests that, in certain contexts, non-commutativity could act as an alternative to the cosmological constant \( \Lambda \), offering a distinct way of modeling the universe's expansion in its early stages.

\section*{Acknowledgments}
Y. Soncco Apaza thanks Fundação de Amparo à Pesquisa do Estado de Minas Gerais (FAPEMIG) for his scholarship. G. Oliveira-Neto also thanks FAPEMIG (APQ-06640-24) for partial financial support. The authors also thank the Federal University of Juiz de Fora (UFJF) for providing the academic environment and facilities that made this research possible.

\end{document}